\documentclass[12pt]{article}

\def\be{\begin{equation}}
\def\ee{\end{equation}}
\def\bse{\begin{subequation}}
\def\ese{\end{subequation}}
\def\ba{\begin{eqnarray}}
\def\ea{\end{eqnarray}}

\usepackage{color,graphicx}
\usepackage{epstopdf,fullpage}
\usepackage{tabularx}
\usepackage{amsmath,amssymb}
\usepackage{hyperref}
\usepackage{natbib}
\newcommand{\KWmm}{K~(W~m$^{-2}$)$^{-1}$}
\newcommand{\coo}{CO$_2$}

\newcommand{\R}{\mathbb{R}}
\newcommand{\Prob}{\mathrm{Prob}}

\begin{document}

\title{State-dependence of climate sensitivity: attractor constraints and palaeoclimate regimes}

\author{Anna S. von der Heydt\thanks{Institute for Marine and Atmospheric Research, Utrecht and Center for Extreme Matter and Emergent Phenomena, Utrecht University, Utrecht, the Netherlands.} \thanks{Correspondence: (A.S.vonderHeydt@uu.nl)} \and Peter Ashwin\thanks{Centre for Systems, Dynamics and Control, Department of Mathematics, University of Exeter, Exeter EX4 4QF, UK.}}

\maketitle

\begin{abstract}
Equilibrium climate sensitivity (ECS) is a key predictor of climate change. However, it is not very well constrained, either by climate models or by observational data. The reasons for this include strong internal variability and forcing on many time scales. In practise this means that the ``equilibrium'' will only be relative to fixing the {\it slow} feedback processes before comparing palaeoclimate sensitivity estimates with estimates from model simulations. In addition, information from the late Pleistocene ice age cycles indicates that the climate cycles between cold and warm regimes, and the climate sensitivity varies considerably between regime because of {\it fast} feedback processes changing relative strength and time scales over one cycle. 
	
In this paper we consider climate sensitivity for quite general climate dynamics. Using a conceptual Earth system model of Gildor and Tziperman (2001) (with Milankovich forcing and dynamical ocean biogeochemistry) we explore various ways of quantifying the state-dependence of climate sensitivity from unperturbed and perturbed model time series.  Even without considering any perturbations, we suggest that climate sensitivity can be usefully thought of as a distribution that quantifies variability within the ``climate attractor'' and where there is a strong dependence on climate state and more specificially on the ``climate regime'' where fast processes are approximately in equilibrium. We also consider perturbations by instantaneous doubling of \coo\ and similarly find a strong dependence on the climate state using our approach.
\end{abstract}


\section{Introduction}

In order to estimate the anthropogenic impact on climate in the future, the response of the climate system to the present perturbation by greenhouse gases needs to be quantified. A frequently used measure for the response to changes in atmospheric \coo-concentration is the equilibrium climate sensitivity (ECS). This is defined as the increase in global mean surface temperature per radiative forcing change after the fast-acting feedback processes in the Earth System have come into equilibrium \citep{Charney1979b}\footnote{In the IPCC literature ECS is frequently given as temperature increase per \coo-doubling, i.e. in units of K, while the equilibrium climate sensitivity parameter describes the warming per radiative forcing, i.e. in units of \KWmm. Here we use the term ECS for both quantities and mostly use the units of warming per radiative forcing.}. Frequently it is estimated by climate model simulations, where the atmospheric \coo-concentration is doubled within a few decades, and equilibrium is assumed after typically 100-200 years; slow climate processes are kept stationary (and non-dynamic) in these model simulations. ECS is the benchmark quantity for climate models\footnote{In \citep{IPCC2013}, regression of the temperature change versus net radiative imbalance at the top of the atmosphere \citep{Gregory2004} is used to estimate ECS.} but is still characterised by a considerable uncertainty of 1.5 -- 4.5~K per \coo-doubling \citep{IPCC2013} and neither recent observations have narrowed down the range of expected climate change  \citep{Knutti2008}. It is also clear that the feedbacks, and hence the ECS will depend on climate state, e.g., \cite{Senior2000,Gregory2004,Crucifix2006,Andrews2008,Yoshimori2011,Caballero2013,Heydt2014}. Palaeoclimate studies have tried to use proxy records to independently constrain ECS \citep{Rohling_palaeosens2012}, but even by taking account of fast feedback processes that depend on the climate state \cite{Heydt2014,Koehler2015} it remains difficult to further constrain the range of expected climate warming. In particular, large temperature changes as a consequence of atmospheric \coo\ increase cannot be excluded. 

The observed warming of the Earth involves both direct radiative forcing and a variety of (positive and negative) fast feedback mechanisms. These are mostly related to atmospheric water vapour content, sea-ice and cloud albedo and aerosol concentrations. On the slower (decadal) time scales ocean heat uptake also contributes to the radiative (im-)balance. Quantifying the (fast) forcing is therefore not an easy task and limits our ability to reduce the uncertainty on climate sensitivity \citep{Schwartz2012}. More generally, the internal variability of the climate system on many time scales leads to a large uncertainty of the ECS. In fact, the most appropriate definition of ECS in the presence of natural variability and forcing is still under debate. Ghil and co-workers propose a non-autonomous stochastic approach in terms of random dynamical systems \citep{Ghil2013,Chekroun2011}, and suggesting that climate sensitivity corresponds to a derivative of a metric (Wasserstein distance) evaluated for the invariant measure with respect to some parameter that is changed. Other approaches include considering perturbations that are not necessarily in the linear regime: \citep{Dijkstra2015} use a conditional nonlinear optimization approach to define climate sensitivity.

In this paper we discuss climate sensitivity as a property of the climate dynamics projected into the space of forcing $R$ to global mean temperature $T$. In particular, Section~\ref{sec:ecs} discusses ECS for the unperturbed system with slow variability in terms of pairs of points on or near the ``climate attractor''. We relate this to more usual concepts of ECS and use this as a way to discuss state dependence. In particular we discuss ``climate regimes'' such that the sensitivity is well constrained within a regime, but poorly constrained whilst switching between regimes. We also discuss possible perturbations and differentiate those that give return to the same climate attractor from those that do not. In Section \ref{s:model} we explore these ideas using a specific conceptual model of the Earth system \citep{Gildor2001a,Gildor2002}, that includes dynamic \coo\ and is able to simulate glacial-interglacial cycles as relaxation oscillations. In this section the model is also perturbed in various ways to obtain (state-dependent) distributions of climate sensitivity. We conclude in Section \ref{s:concl} with a discussion of some issues that arise related to the extraction and interpretation of ECS distributions from palaeoclimate records.


\section{Climate sensitivity and dynamics in $(T,R)$ space}
\label{sec:ecs}

The usual approach to ECS is to consider the radiative energy balance for the global mean surface temperature $T$
\begin{equation}
\frac{d T}{d t} \sim R_{f}  + R_{slow} + R_{fast} - R_{OLW},
\label{e:energybalance}
\end{equation}
where $R_{f}$ we understand as the radiation due to (external) forcings, e.g. the radiation received from the sun $R_{ins}$ and the forcing $R_{{\rm CO}_2}$ due to the greenhouse effect of atmospheric \coo. The fluxes $R_{slow}$ and $R_{fast}$ are contributions to the radiative balance due to a set of feedback processes in the climate system that are classified as slow and fast, respectively, usually relative to the typical time scale of the forcing. Finally, $R_{OLW}=-\varepsilon\sigma_B T^4$ is the outgoing longwave radiation determined by the Stefan-Boltzmann law (with $\sigma_B$ the Stefan-Boltzmann constant and $\varepsilon$ the emissivity of the atmosphere). The equation (\ref{e:energybalance}) suggests a simple $(T,R)$ relationship but in reality it is hard to unpick this relationship, not least because fast feedbacks may potentially give multiple attractors when the slow feedbacks are fixed.

Here, we simply take the approach that $T$ and $R$ are quantities that one can (in principle) observe from a complex system and we investigate the relationship between them. Given two climate states with forcings $R_{f,1}$ and $R_{f,2}=R_{f,1}+\Delta R$ and temperatures $T_{1}$ and $T_{2}=T_{1}+\Delta T$, we consider climate sensitivity as the ratio of temperature change to radiative forcing change (including slow feedbacks):
\be
S = \frac{T_{2}-T_{1}}{R_{f,2}+R_{slow,2}-R_{f,1}-R_{slow,1}} =  \frac{\Delta T}{\Delta R_{f}+\Delta R_{slow}}.
\label{e:Sspecific}
\ee
If there is a functional relationship of the form $T=T(R)$ for $R=R_{f}+R_{slow}$ then in the limit of small differences in $R$ we expect
\be
S \approx \frac{dT}{dR}
\label{e:Sslope}
\ee
In the (hypothetical) case that all slow processes are known and quantified and that a time-scale separation between slow and fast processes exists, $S$ is the usual ECS. However, we take the approach that (\ref{e:Sspecific}) can still be studied when no functional relation $T(R)$ exists, and this naturally leads to distributions of ECS.

If we consider only \coo\ as forcing and the land-ice albedo feedback as slow process (i.e. $R_f+R_{slow,1}=R_{[CO_2,LI]}$) then the specific climate sensitivity is the ratio
\begin{equation}
S_{[CO_2,LI]} = \frac{T_2-T_1}{R_{[CO_2,LI],2}-R_{[CO_2,LI],1}}= \frac{\Delta T}{\Delta R_{[CO_2,LI]}}.
\label{e:Sapprox}
\end{equation}
{\em A priori} it is not clear how (\ref{e:Sapprox}) depends on the two climate states being compared. Moreover, on the one hand $\Delta R$ should be small to make a linear approximation valid. On the other hand, taking climate states where $\Delta R$ is large is more likely to give values that are insensitive to measurement errors, in particular for palaeoclimate records. Indeed, $S_{[CO_2,LI]}$ will only give a single value if $T$ is a (smooth) function of $R_{[CO_2,LI]}$ and we consider asymptotically small $\Delta R$: in this case the distribution approaches a $\delta$ function centred at $dT/dR$.

\subsection{Climate sensitivity on the climate attractor}

Let us suppose that there is an attractor for the climate system that is stationary (this includes the possibility of a climate that is turbulent and/or that responds in a chaotic way to stationary quasi-periodic astronomical forcing), and that the system is on a trajectory that explores this attractor as time progresses. Comparing the climate states at times $t_{ref}$ and $t_{ref}+\delta$, one can define climate sensitivity over a time interval $[t_{ref},\delta]$ as
\be
S_{[CO_2,LI]}(t_{ref},\delta) = \frac{T(t_{ref}+\delta)-T(t_{ref})}{R_{[CO_2]}(t_{ref}+\delta)+R_{[LI]}(t_{ref}+\delta)-R_{[CO_2]}(t_{ref})+R_{[LI]}(t_{ref})}.
\label{e:Sdelta}
\ee
By considering a range of possible reference times $t_{ref}$ and delays $\delta$ there will be a distribution of sensitivities. 

An alternative approach is to assume there is a stationary measure (or distribution) $\mu$ of points in the $(T,R_{[CO_2,LI]})$-plane weighted according to how often they are visited over asymptotically long times, i.e. we assume that 
\begin{equation}
\mu(A):=\lim_{t'\rightarrow \infty} \frac{1}{t'}\ell\left(\{0<s<t'~:~(T(t+s),R_{[CO_2,LI]}(t+s))\in A\}\right).
\label{e:freqTvsR}
\end{equation}
is independent of $t$ and typical initial condition, where $\ell(B)$ is the length of the set $B\subset \R$: see Figure~\ref{f:schematicTR}a. The measure $\mu$ can be thought of as a projection of a natural measure on the attractor onto the two observables $(T,R_{[CO_2,LI]})$: it gives a distribution of associations between $T$ and $R_{[CO_2,LI]}$.

This distribution\footnote{This can be thought of as projection of an invariant measure on a climate attractor \citep{Chekroun2011} onto these observables.} naturally leads to a distribution of climate sensitivities by picking pairs of points $(R_{[CO_2,LI] 1,2},T_{1,2})$ that are independently distributed according to $\mu$ and evaluating (\ref{e:Sapprox}). In other words, for any (measurable) $A\subset \R$ we can use $\mu$ to assign a probability to the sensitivity being in $A$:
\be
\Prob(S_{[CO_2,LI]}\in A):=\mu\times\mu\left(\left\{ (T_1,R_{[CO_2,LI],1}),(T_2,R_{[CO_2,LI],2})~:~S_{[CO_2,LI]} \in A\right\}\right).
\label{e:Sdist}
\ee
Note that (\ref{e:Sdelta}) can be determined from a time series that does not necessarily explore the full attractor, while (\ref{e:Sdist}) considers states purely depending on the locations in the $(T,R)$ plane. In section 3, we give an example showing that the approaches (\ref{e:Sdelta}) and (\ref{e:Sdist}) can give similar distributions when considering a wide range of $\delta$ and initial points.

\subsection{Climate sensitivity, regimes and responses to perturbation}

A study of palaeoclimate records, for example the ice age cycles of the last 800 kyr, shows the presence of markedly different ``regimes'' of climate, namely periods of slowly varying climate and rapid transitions between these regimes (the deglaciations). We wish to evaluate the sensitivities associated within one regime, and associated with changing regimes. For definiteness, we only consider climates with two regimes - a cold ($C$) and a warm ($W$) regime - in this paper. If one partitions the attractor into two regimes in state space, this implies a partition of $\mu$ into two distributions
$$
\mu=\mu_C+\mu_W.
$$
Evaluating the distribution of climate sensitivities corresponding to choosing typical endpoints relative to these distributions allows one to examine the sensitivities within regimes. In particular one can define conditional distributions of sensitivities of the ``warm'' (and similarly the ``cold'') states by
\be
\Prob(S^{WW}_{[CO_2,LI]}\in A):=\mu_W\times\mu_W\left(\left\{ (T_1,R_{[CO_2,LI],1}),(T_2,R_{[CO_2,LI],2})~:~S_{[CO_2,LI]} \in A\right\}\right)
\label{e:SWW}
\ee
where $S_{[CO_2,LI]}$ is as in (\ref{e:Sapprox}). 
There are conditional sensitivities associated with regime changes, for example from $C$ to $W$ this is
\be
\Prob(S^{CW}_{[CO_2,LI]}\in A):=\mu_C\times\mu_W\left(\left\{ (T_1,R_{[CO_2,LI],1},T_2,R_{[CO_2,LI],2})~:~S_{[CO_2,LI]} \in A\right\}\right)
\label{e:SCW}
\ee
and the distribution (\ref{e:Sdist}) can be thought of as the sum of the conditional distributions for $S^{WW}$, $S^{CC}$, $S^{CW}$ and $S^{WC}$. Note that from the definition above
$$
\Prob(S^{CW}_{[CO_2,LI]}\in A)=\Prob(S^{WC}_{[CO_2,LI]}\in A),
$$
even if physically and when time progresses the $CW$ transition is different than the $WC$ transition.
For an optimal choice of regimes one would aim to ensure that the distribution of sensitivities within each regime is tightly localised, while those associated with regime changes may be poorly localised. A regime could, therefore, be defined as a region in $(T,R)$ space where the $T(R)$ relation is almost linear.

\begin{figure}[ht!]
	\centerline{\includegraphics[width=15cm]{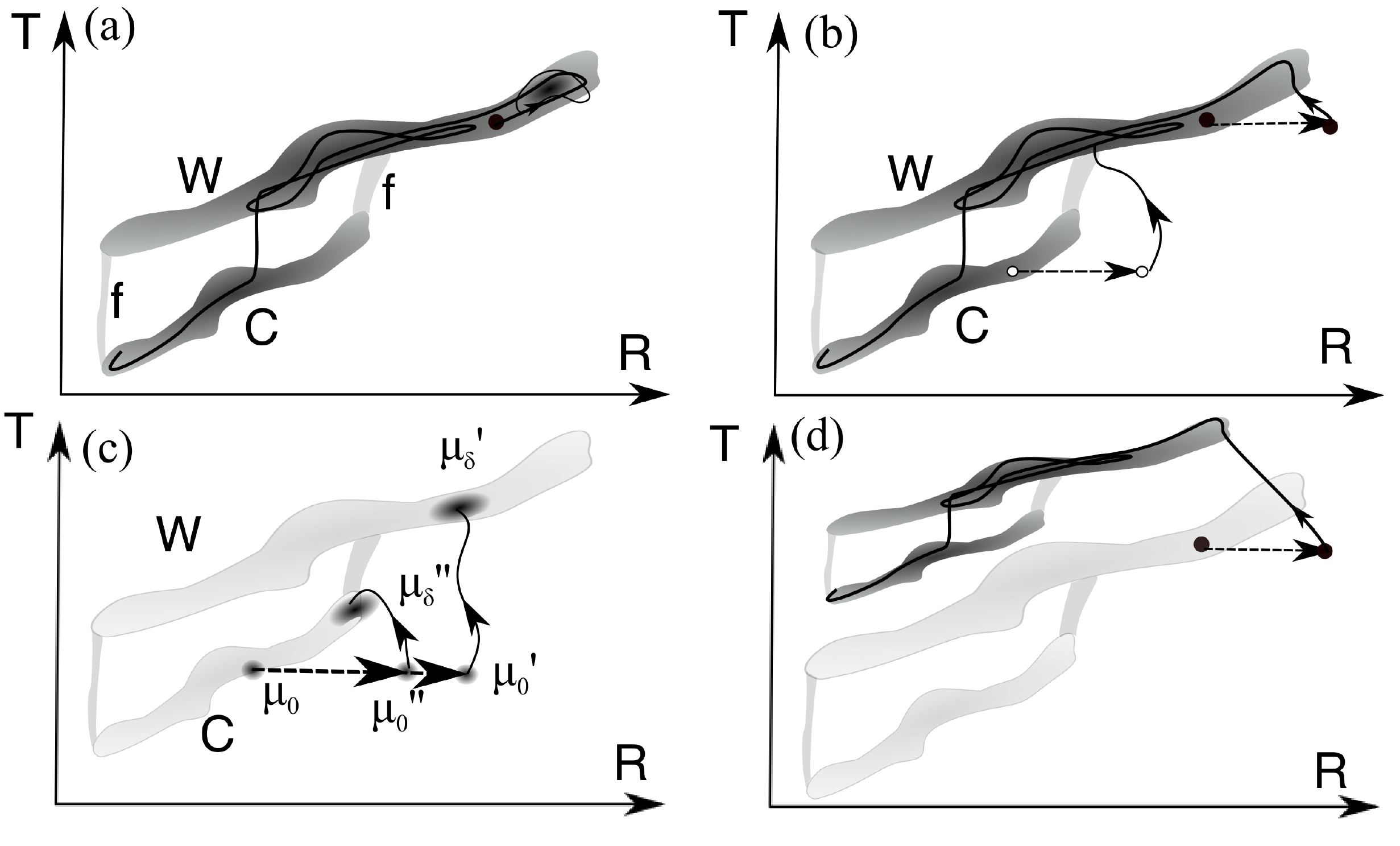}}
	\caption{Schematic diagram showing global mean temperature $T$ versus radiative forcing $R$ due to atmospheric \coo.
    (a) In the presence of natural forcing we assume there is a stationary distribution $\mu$ (shown by grey scale) in the $(T,R)$-plane - this is the projection of a dynamical measure onto this plane and can be divided into two climate regimes for the slow dynamics (shown here as $C$ and $W$ states), linked by fast changes (shown as $f$). Picking two points relative to this measure gives a distribution of slopes that quantifies long-term variability of climate sensitivity for this forcing.
    (b) A small impulsive change of $R$ that does not structurally change the system (dashed line) takes the system state away from the attractor. If the perturbation does not change the attractor, after a transient (small arrow), we expect to continue to explore the plane according to the distribution $\mu$. Depending on where the perturbation is applied and its size, the response may involve a switch between different regimes of the attractor (see the perturbation applied to $C$ state). 
        (c) A small impulsive change $R$ (dashed line) moves an initial distribution $\mu_0$ to a new location $\mu_0'$ or $\mu_0''$ away from the attractor. After some time $\delta>0$ we reach a perturbed distribution $\mu_{\delta}'$ or $\mu_{\delta}''$: these may be in different regimes depending on the initial state and strength of the perturbation.
    (d) A large, or structural change to the system will give a new attractor and a different set of asymptotic states.
    }
	\label{f:schematicTR}
\end{figure}

We now consider response to two types of instantaneous perturbation. The first type of perturbation does not structurally change the system or leave the basin of the current attractor: the response shows transient decay back to the attractor followed by continued motion on the same attractor. The response to such a perturbation includes the possibility of switching between regimes of the attractor, depending on the initial point on the attractor where the perturbation is applied. Figure~\ref{f:schematicTR}b illustrates this schematically. In such a case, the distribution of sensitivities will potentially depend on the timescale $\delta$ of interest and the initial time $t_{ref}$. However we expect it to decay to the (regime dependent) sensitivity for large $\delta$. 

The second type of perturbation either structurally changes the system attractor, or is large enough to place the state in the basin of a different attractor. In either case the response will approach a new attractor as illustrated in Figure~\ref{f:schematicTR}d. In this case the distribution of sensitivities obtained by comparing initial and final states may not resemble regimes of either attractor.

Finally, we mention another approach to perturbation that is particularly useful for short-term prediction: Figure~\ref{f:schematicTR}c starts with a localised (say Gaussian) distribution $\mu_0$ centred on a perturbed reference state at some time $t_{ref}$, and propagate this forwards to time $t_{ref}+\delta$ for some $\delta>0$. The initial distribution will spread to give a localised measure $\mu_{\delta}$ that gives a distribution of possible sensitivities via (\ref{e:Sapprox}), which again may depend on the timescale $\delta$, which again may depend on the timescale $\delta$. ECS derived from palaeoclimate records typically reflects the climate sensitivity on the attractor (Fig.~\ref{f:schematicTR}a), while model-determined ECS usually involves some type of perturbation (away from the attractor). In this sense, palaeo ECS and model ECS are conceptually different. In the next section, we explore how and when these two concepts can still give similar disributions, using a conceptual Earth System model.


\section{Climate sensitivity in a conceptual climate model}
\label{s:model}

The conceptual model of the climate system of \cite{Gildor2001a, Gildor2002} has been shown to simulate the glacial-interglacial transitions; the model equations are given in Appendix~\ref{ap:model}. In this model, the atmosphere is represented by 4 meridional boxes while the ocean component consists of two layers of 4 meridional boxes each. The model includes land ice, sea ice and carbon-cycle effects, such that the atmospheric \coo\ concentration is a dynamic variable in the model. The model contains one dynamic fast feedback, namely the sea ice-albedo feedback evolving on (sub-)decadal time scales, and one slow feedback, the land ice-albedo feedback, which evolves on the order of millennial time scales. On the decade-to-century time scale the model includes an additional process in the surface radiative balance due to heat exchange between ocean and atmosphere. All other fast feedbacks (water vapor, clouds, aerosols, lapse rate) are represented by a fixed temperature response to the radiative forcing in the system. In this case, as discussed in Appendix~\ref{ap:cs} there is only one active slow feedback process, and the specific climate sensitivity parameter $S_{[CO_2,LI]}$ (\ref{e:Sapprox}) represents the model's ECS. Orbital forcing is included in the model through varying incoming solar radiation averaged over each atmospheric box on seasonal and orbital time scales and modulating the Northern Hemisphere land ice ablation term by the (northern polar box averaged) summer insolation on orbital time scales \citep{Gildor2000}. 

The atmospheric \coo-concentration in the full model system deserves further discussion because it can be viewed as both a forcing and a feedback. While the dynamic \coo\ is not essential for generating the glacial-interglacial cycles in the model, it feeds back on their amplitude; during cold periods when land ice is growing, reduced vertical mixing in the Southern Ocean and extended Southern Ocean sea ice cover leads to reduced atmospheric \coo\ \citep{Gildor2002}. The exchange of \coo\ between ocean and atmosphere is fast (on time scales of a decade), however, the vertical mixing of surface-to-deep water masses in the Southern Ocean is affected by the temperature of the North Atlantic deep water, which evolves on slower time scales. The associated feedback process therefore acts on various time scales. 
When determining climate sensitivity, \coo\ is generally assumed a forcing. Here, it is important to keep in mind that it also can be viewed as a feedback as has been observed also in other models \citep{Scheffer2006} and will be the case in the real climate system, particularly on long (geological) timescales. 

\subsection{Glacial-interglacial cycles as relaxation oscillations}
\label{ss:glacialcycles}
\begin{figure}[ht!]
\centerline{\includegraphics[width=12cm]{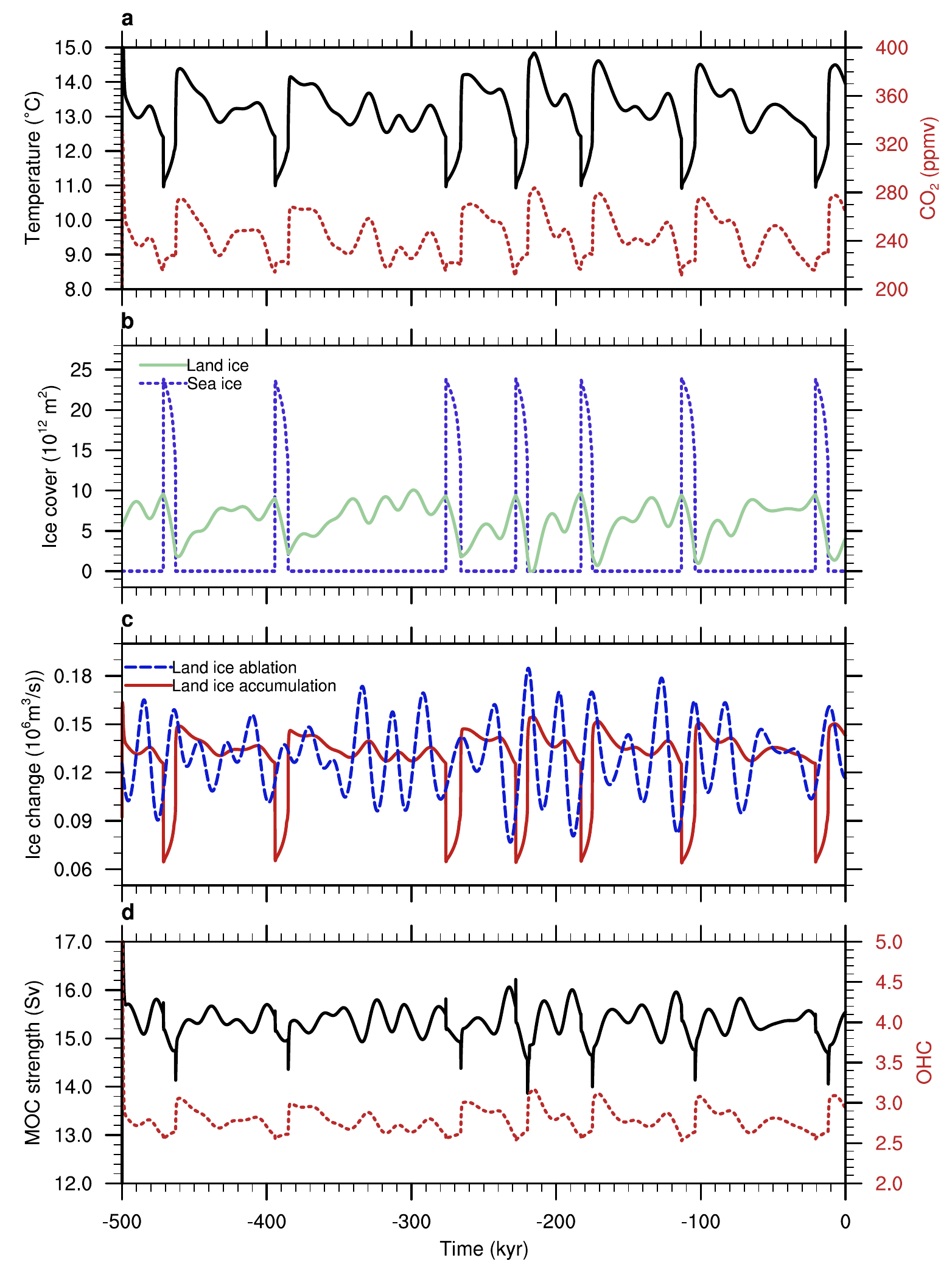}}
\caption{Glacial cycles of the box model, shown are time series of 100-year averages.
(a) Simulated global mean surface temperature $T$ (black line) and atmospheric \coo\ (red line); 
(b) Land ice (green line) and sea ice (blue line) cover of the northern polar box; 
(c) Land ice accumulation (red line) and ablation (blue line) in the northern polar box; 
(d) Strength of the ocean meridional overturning circulation (black line), measured as the volume exchange between surface and deep northern polar ocean boxes and ocean heat content (red line). 
}
\label{f:glacialcycles}
\end{figure}

We first analyse a simulation with the climate model including prognostic $pCO_2$ and Milankovitch forcing. The simulation is started 500 kyr ago from initial conditions that are assumed to be close to the attractor and it is run up until the present day. The simulated glacial-interglacial cycles show a peak-to-peak global mean temperature difference of up to 4~K (Fig.~\ref{f:glacialcycles}a). Corresponding CO$_2$ differences are about 75 ppmv, which here are completely generated by the effect of the solubility pump in the ocean.

\begin{figure}[ht!]
\center\includegraphics[width=7cm]{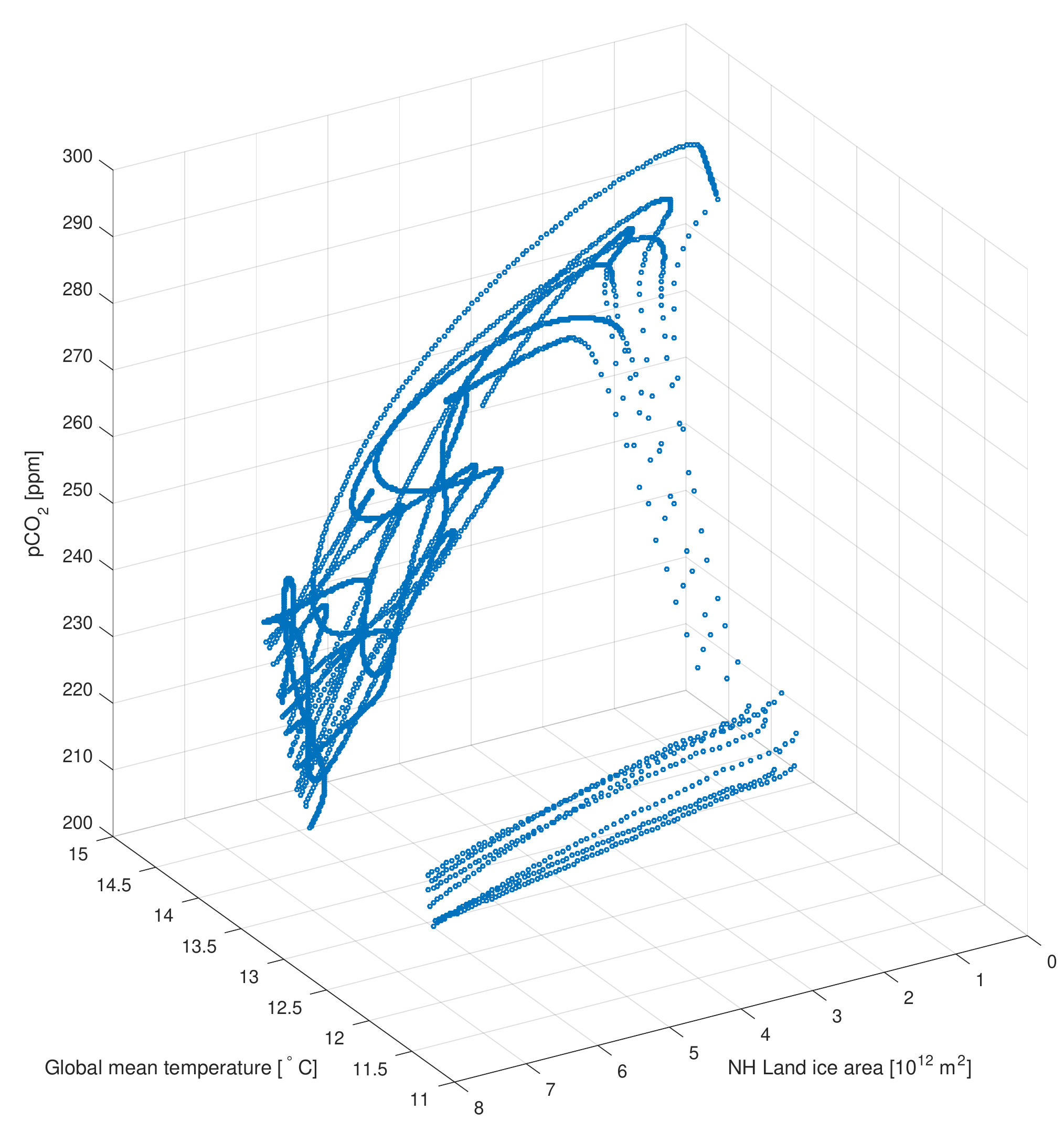}~\includegraphics[width=7cm]{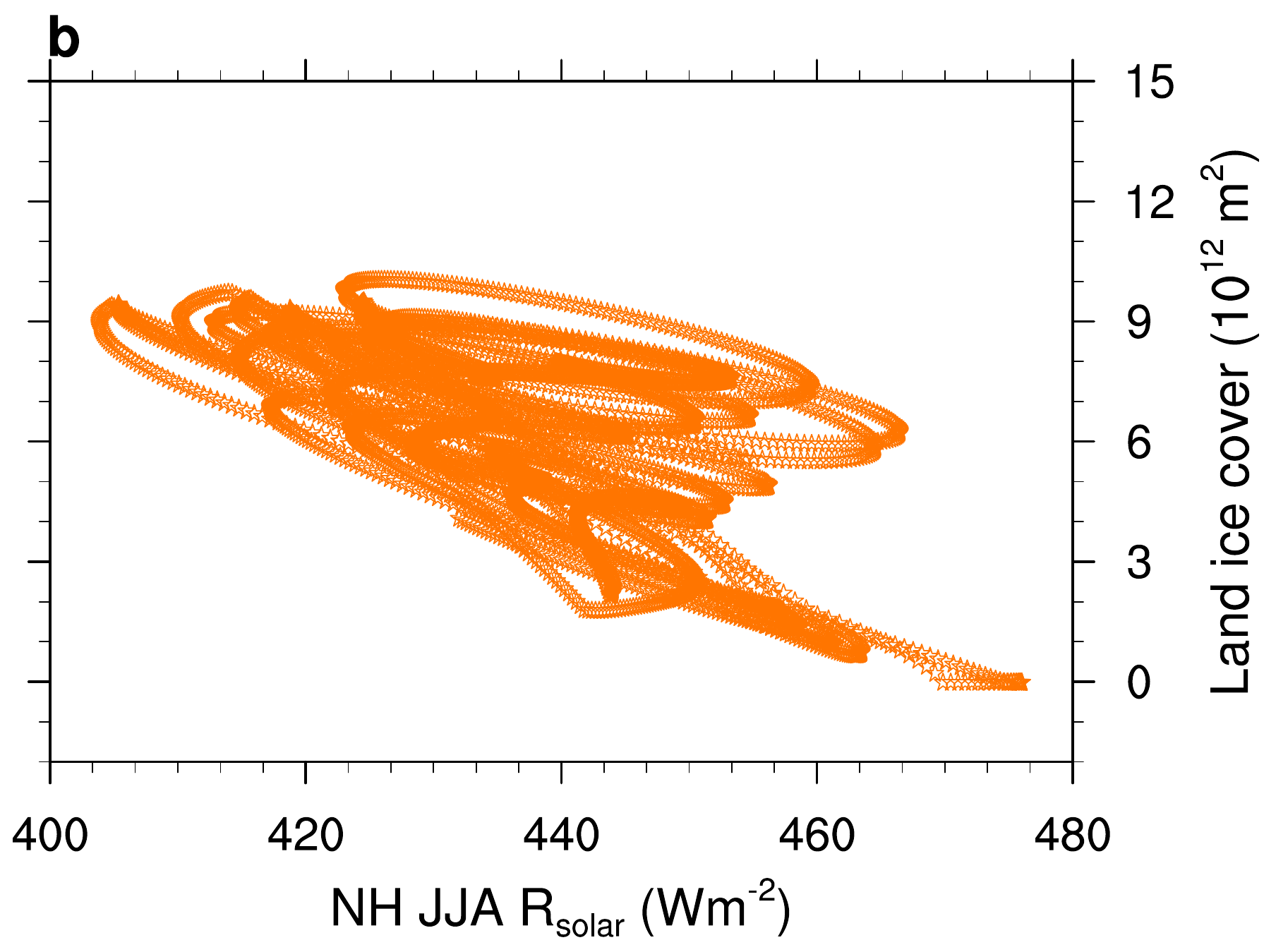}
\caption{Phase diagrams of the glacial cycles of the box model; each point is a 100-year average.
(a) NH land ice cover as a function of global mean temperature and atmospheric \coo; 
(b) NH land ice cover as a function of orbital variations in solar radiation $R_{solar}$ defined as summer insolation (JJA) over the northern polar box of the model (45-90$^\circ$ averaged).
}
\label{f:glacialcycles_p}
\end{figure}

In this model the fast sea ice-albedo feedback is responsible for the abrupt glacial-interglacial variations --- the so-called sea ice switch mechanism as suggested by \citet{Gildor2001a}.  The sea ice switch mechanism generates the glacial cycles in the model as self sustained relaxation oscillations because the ice volume thresholds for switching sea ice cover `on' and `off' differ \citep{Crucifix2012a} (see Fig.~\ref{f:glacialcycles_p}a) and we use this to define two climate regimes, a cold $C$ regime with extensive sea ice cover and a warm $W$ regime without sea ice. When the land ice volume slowly grows (accumulation exceeds ablation), the atmospheric and surface ocean temperature decrease due to increasing albedo of the planet.

Once the polar surface ocean temperature has reached a critical value cold enough to form sea ice, the polar box is rapidly covered with sea ice, which further reduces the atmospheric temperature through the ice-albedo feedback and prevents evaporation from the polar ocean box (Fig.~\ref{f:glacialcycles}b). In addition, atmospheric moisture content is reduced due to lower temperatures, which leads to decreasing land ice volume (accumulation is smaller than ablation, Fig.~\ref{f:glacialcycles}c). Temperature starts rising again both due to smaller albedo and because the ocean warms below the insulating sea-ice cover until it is warm enough to melt the polar sea ice, Fig.~\ref{f:glacialcycles}d). At this point there is a change in regime: the global temperature quickly rises, moisture content in the atmosphere increases and the land ice starts growing again (accumulation becomes larger than ablation). 

In this model, Milankovitch forcing is not necessary to generate the glacial-interglacial cycles, but it modifies them and makes them more irregular. Although there is some degree of synchronisation of the glaciation and deglaciations to the orbital forcing, the relation between land ice and global mean solar radiation is not trivial (Fig.~\ref{f:glacialcycles_p}b). Milankovitch forcing mainly modulates the (otherwise constant) ablation of the northern hemisphere land ice, and therefore, while the land ice is growing and ocean temperatures decreasing in some (slightly warmer) periods, land ice accumulation becomes smaller than ablation and the ice growth and ocean cooling is reversed for a while (Fig.~\ref{f:glacialcycles}c).

\subsection{Climate sensitivity for unperturbed climates}
\label{s:model_long}

\begin{figure}[ht!]
\center\includegraphics[width=10cm,clip=]{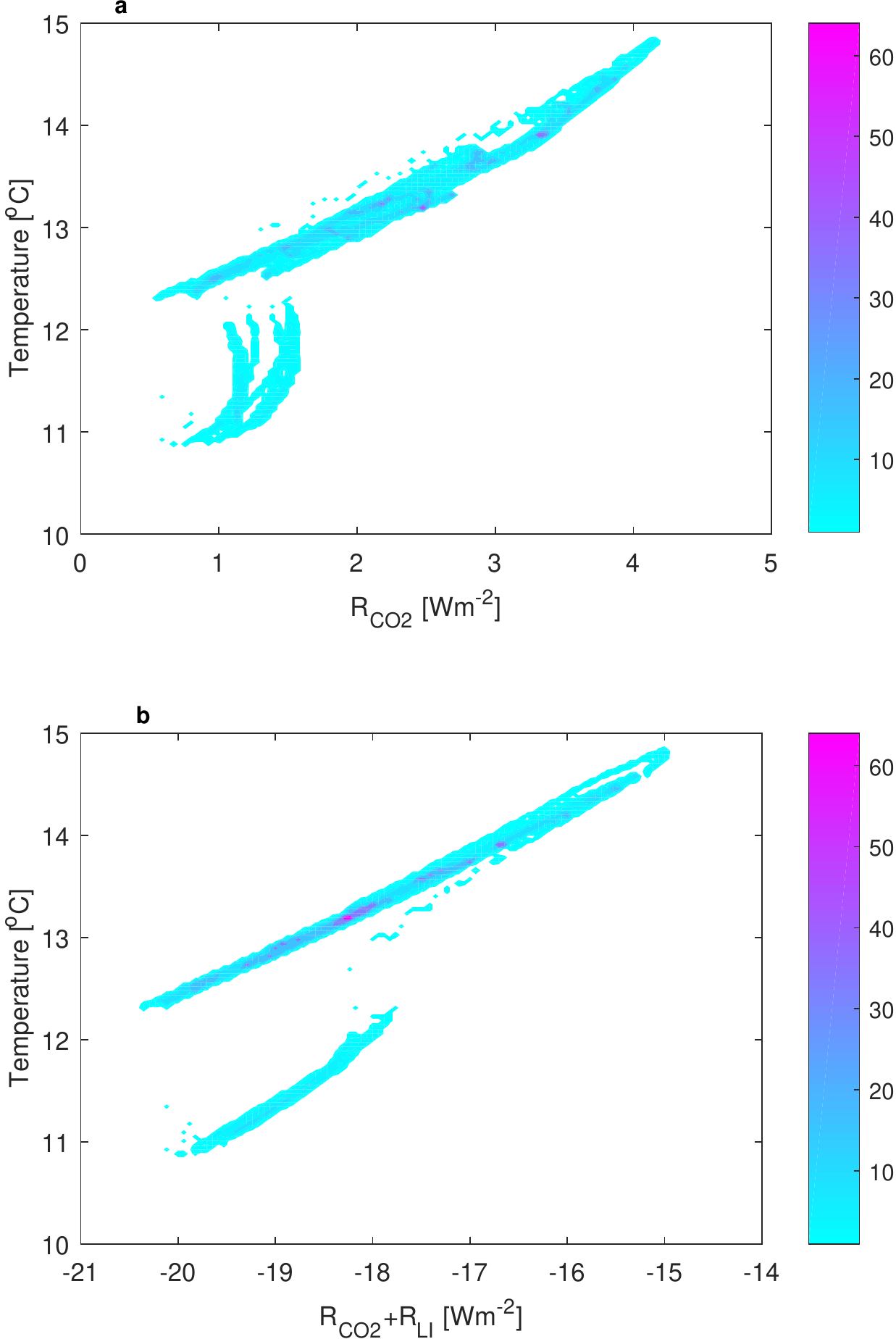}
\caption{Relation between global mean temperature $T$ and radiative contributions $R$ due to pCO$_2$ $R_{[CO_2]}$ and land ice  $R_{[LI]}$ (slow feedback). Colours [a.u.] show a box-counting approximation of the probability density of distribution of the relation, after a transient has been removed (see text for details):
	(a) distribution of $T$ versus $R_{CO_2}$;
	(b) distribution of $T$ versus $R_{CO_2,LI}$. 
	Observe the presence of two clear regions of high density: an upper regime of $W$= warm states and a lower regime of $C$=cold states where there is extensive sea ice present. Note the fast switches $f$ between state contain very little density as they are much faster than the evolution within the $W$ and $C$ states.}
\label{f:localdensity}
\end{figure}

If one tries to determine climate sensitivity from past climate records, there is only one temporal realisation of a trajectory on the climate attractor that can be measured: perturbations away from the attractor cannot be performed. Defining the climate sensitivity in terms of the measure on the climate attractor (see Fig.~\ref{f:schematicTR}a), we need to consider the relation between temperature $T$ and radiative forcing due to \coo\ and land ice (the only slow process in the climate model). Fig.~\ref{f:localdensity}a,b shows the probability density of $(T,R)$ combinations for the 500 kyr trajectory discussed above, obtained by box-counting the frequency of visits to a uniform discretization of this range of $T$ and $R$ into $120\times 120$ cells (we remove a transient of length 10,000 yrs). This empirical distribution can be seen as an approximation of $\mu$ in (\ref{e:freqTvsR}).

In the special case of this relation being a linear function $T(R)$, the climate sensitivity is given by the slope of this line and constant for all climate states. However, it has been previously shown that in this climate model the climate sensitivity is strongly state dependent due to the fast sea-ice albedo feedback changing in strength between different climate states \citep{Heydt2014}, which allows the definition of a local climate sensitivity (for a reference climate state). Indeed, in Fig.~\ref{f:localdensity}b there appear to be regimes where the $(T,R)$ relation is close to a (linear) function, but particularly in the transition region from glacial to interglacial states the sensitivity is less well defined or even negative. 

From this data we first estimate the slope of the relation $S_{local}$ between $T$ and $R_{[CO_2,LI]} = R_{[CO_2]}+R_{[LI]}$ (Fig.~\ref{f:localdensity}b) by a linear regression on the warm ($W$) and cold ($C$) parts of the data, giving $S^W_{local} = 0.45$\KWmm, and $S^C_{local} = 0.54$\KWmm, respectively. In the regression all points with $T\le 12^\circ$C are considered for $S^C_{local}$ and points with $12.5^\circ$C$\le T\le 14.5^\circ$C for $S^W_{local}$, respectively. Note that the temperature classification divides the data into climate states without northern hemisphere sea ice ($W$) and those where sea ice is present ($C$). The physical explanation for the higher sensitivity during the colder part of the data is that the presence of sea-ice in the cold climate states leads to a stronger sea ice-albedo feedback. 

\begin{figure}[ht!]
\center\includegraphics[width=15cm]{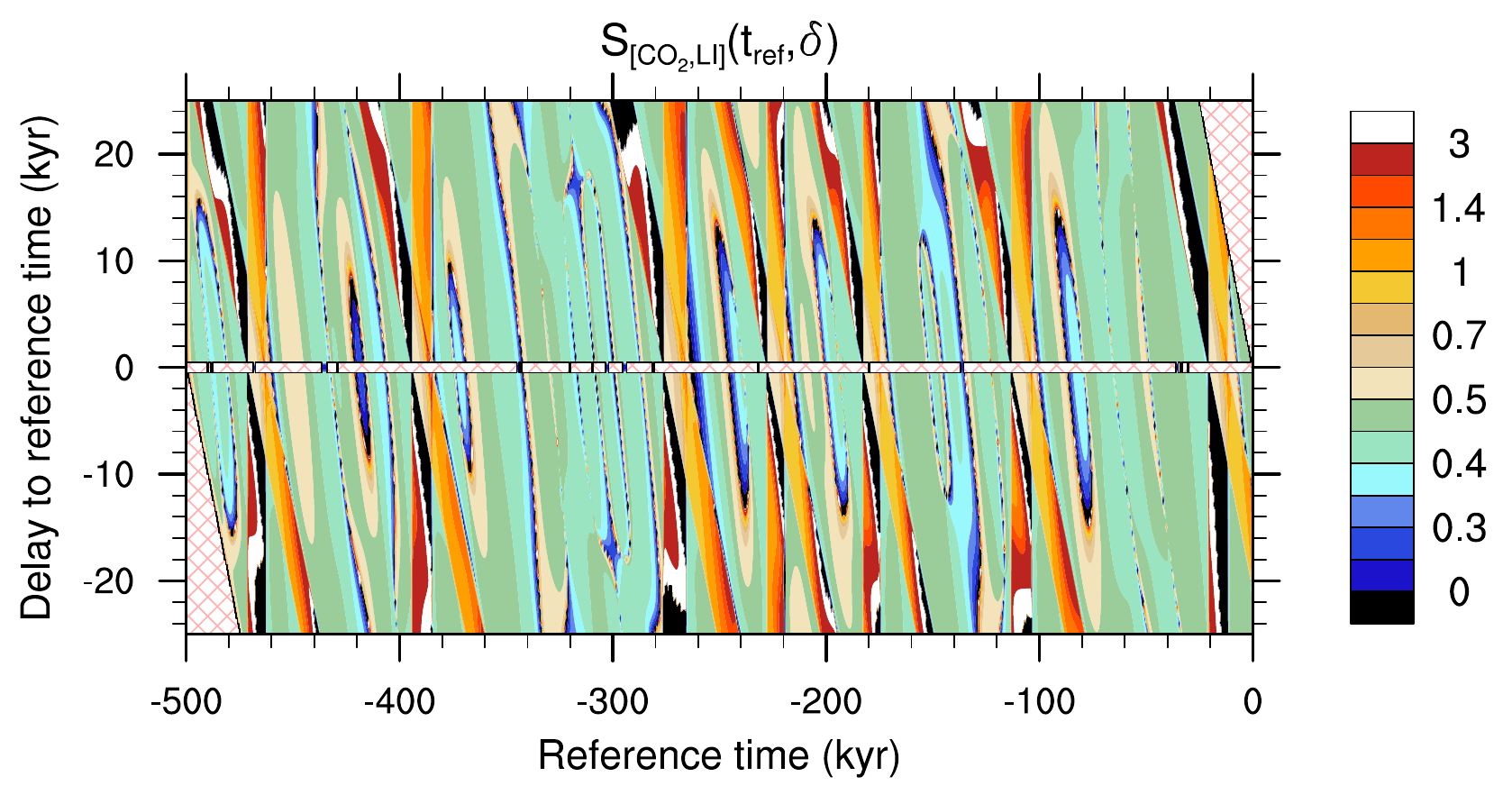}
\caption{Equilibrium climate sensitivity approximated by the specific climate sensitivity $S_{[CO_2,LI]}(\delta,t_{ref})$ from equation (\ref{e:Sdelta}) from a long glacial-interglacial simulation. All reference times along the time series are considered and delays to the reference time between $\delta = -25 .... +25$~kyr (about half a glacial-interglacial period). The plot shows contours of $S_{[CO_2,LI]}(\delta,t_{ref})$ as a function of $\delta$ and $t_{ref}$. White shading indicates values of $S_{[CO_2,LI]}(\delta,t_{ref})$ above 3~\KWmm, black shading indicates negative values.  
}
\label{f:Sdeltref}
\end{figure}

\begin{figure}[ht!]
\center\includegraphics[width=14cm]{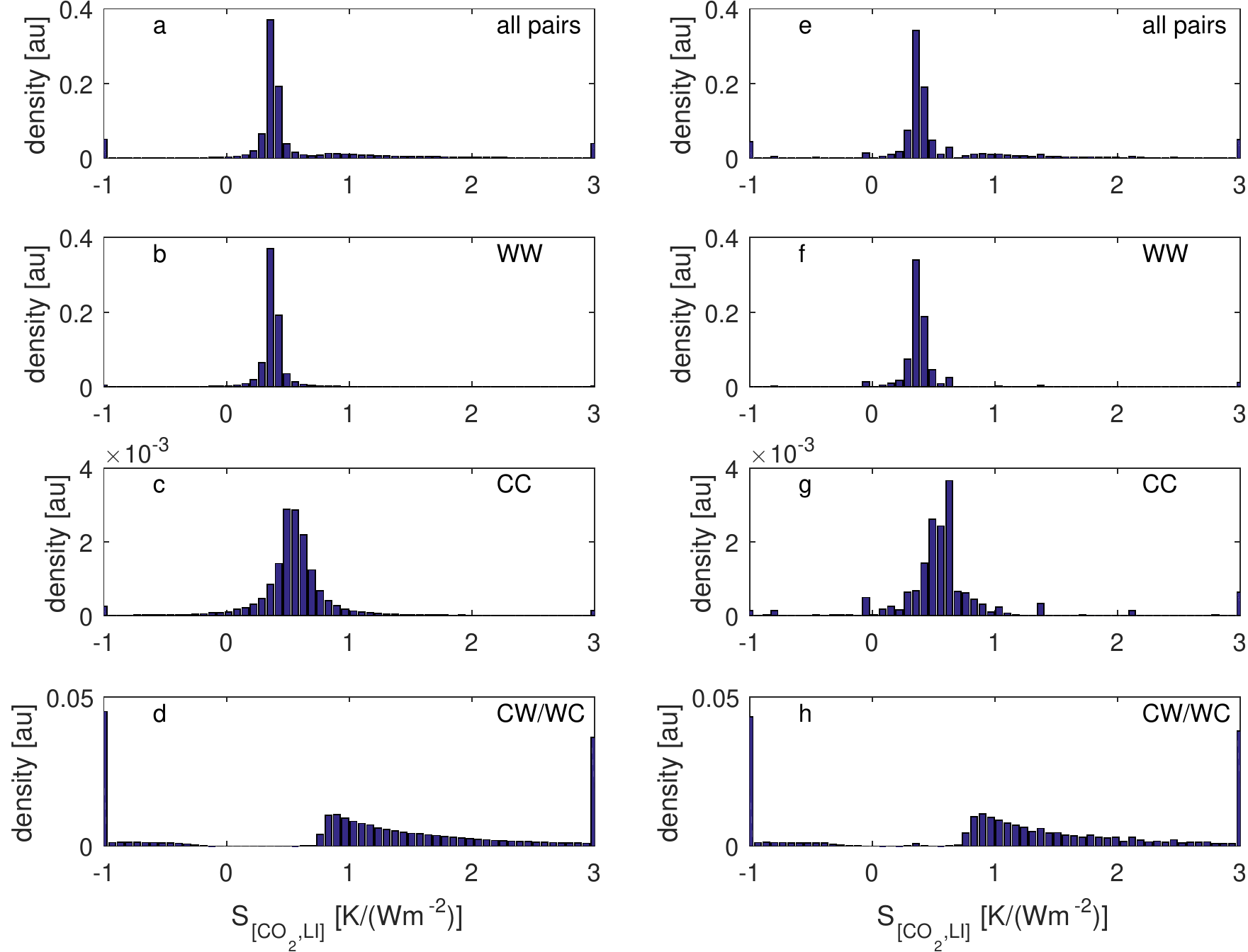}

\caption{Distributions of the specific climate sensitivity $S_{[CO_2,LI]}$. The {\bf left panels} (a-d) use (\ref{e:Sdelta}) and a long glacial-interglacial simulation for a range of reference times and a range of delays greater than $500$ yr. The {\bf right panels} (e-h) use (\ref{e:Sapprox}) and the approximation of $\mu$ in Figure~\ref{f:localdensity}b. Values of $S$ outside the range $[-1,3]$ are truncated to the endpoints of the domain.
(a,e) All climate states are considered;
(b,f) Only pairs of climate states that are both in the $W$ regime (no sea ice) are considered;
(c,g) Only pairs of climate states that are both in the $C$ regime (sea ice present) are considered;
(d,h) Only pairs of climate states, where one is in the $W$ regime and the other is in the $C$ regime are considered. 
Observe that within each regime the distribution appears to be fairly tightly defined, whilst the $CW/WC$ transitions have very long tails. Moreover, the distributions from the two methods give comparable results both within and across regimes.}
\label{f:Sclassified}
\end{figure}

As can be seen by the density of points  in Fig.~\ref{f:localdensity}b, even the almost linear parts of the $(T,R)$ relation are not a (smooth) function: there is a distribution of slopes for each climate state. In Fig.~\ref{f:Sdeltref}, $S_{[CO_2,LI]}(\delta,t_{ref})$ is shown for all values of $t_{ref}$ and delays $\delta = 0 - \pm  25$~kyr (1/4 of the average period of the glacial-interglacial cycles), where the white (black) shading indicates very large ($\ge 3$ \KWmm) positive (all negative) values. A distribution of $S_{[CO_2,LI]}(\delta,t_{ref})$ is given in Fig.~\ref{f:Sclassified}a. 
We also classify $S_{[CO_2,LI]}(\delta,t_{ref})$ in terms of which regimes ($C$ or $W$) are being compared at times $T_{ref}$ and $T_{ref}+\delta$. The resulting distributions  $S^{WW}_{[CO_2,LI]}(\delta,t_{ref})$, $S^{CC}_{[CO_2,LI]}(\delta,t_{ref})$ and $S^{CW/WC}_{[CO_2,LI]}(\delta,t_{ref})$ are shown in Fig.~\ref{f:Sclassified}b--d. Comparing only $W$ states (without sea ice) leads to generally lower sensitivity, with its mean close to the value determined by the linear regression of only $W$ states and a rather narrow distribution. Similarly, comparing only $C$ states (with variable sea ice) results in somewhat higher sensitivities and a larger spread around the mean (which is again close to the linear regression of the $C$ states). The larger spread is not surprising given that the $(T,R)$ relation in the $C$ regime is clearly nonlinear. 

The plots in Fig.~\ref{f:Sclassified}e-h correspond to Fig.~\ref{f:Sclassified}a-d but are calculated using (\ref{e:Sdist}) and the approximate measure $\mu$ whose density is shown in Fig.~\ref{f:localdensity}b. Note the similarity of the distributions found by both methods. The largest (and most negative) values of $S_{[CO_2,LI]}$ in (Fig.~\ref{f:Sclassified}a,e) originate from the cross-comparison of $C$ and $W$ regimes (Fig.~\ref{f:Sclassified}d,h). On the other hand, the means of $S^{CC}_{[CO_2,LI]}$ agrees well with $S^C_{local} = 0.54$\KWmm while the mean of $S^{WW}_{[CO_2,LI]}$ agrees well with $S^W_{local} = 0.45$\KWmm.

This suggests (a) the discretization used to approximate $\mu$ is sufficient to capture the main features of the regime-dependent sensitivity and (b) that the reference times and delays considered sample the distribution $\mu$ well and so distributions of sensitivities given by (\ref{e:Sdist}) and (\ref{e:Sdelta}) are comparable.

\subsection{Climate sensitivity for perturbed model climates}
\label{ss:dynamic_p}
If we consider climate sensitivity as a local property of a natural measure on the climate attractor as illustrated in Fig.~\ref{f:schematicTR}a, we need to explore the set of points in the $(T,R)$-plane that the model visits over the glacial-interglacial cycles. In climate models used for future prediction the usual approach is to perturb the system in some way (e.g. double p\coo) and study the response to this perturbation after some time. An initial distribution $\mu_0$ evolves over a certain time scale $\delta$ after a perturbation away from the attractor has been applied as illustrated in Fig.~\ref{f:schematicTR}b,c. The distribution of sensitivities can be found by perturbing the system instantaneously, assuming that the perturbation is not too large and the system returns to the same attractor after a transient. Note, however, that this approach requires a different type of perturbation than what is usually applied in climate models; the standard procedure in GCMs is to consider a prescribed (non-dynamic) \coo-doubling as a perturbation, where in fact a different attractor than that of the full Earth system (including dynamic carbon cycle) is explored. In this section, we derive the model's ECS in response to a perturbation to the initial atmospheric \coo\ including a dynamic carbon cycle, and evaluating the temperature response to this initial perturbation at different times (following the approach illustrated in Fig.~\ref{f:schematicTR}c). 

\begin{figure}[ht!]
\center\includegraphics[width=12cm]{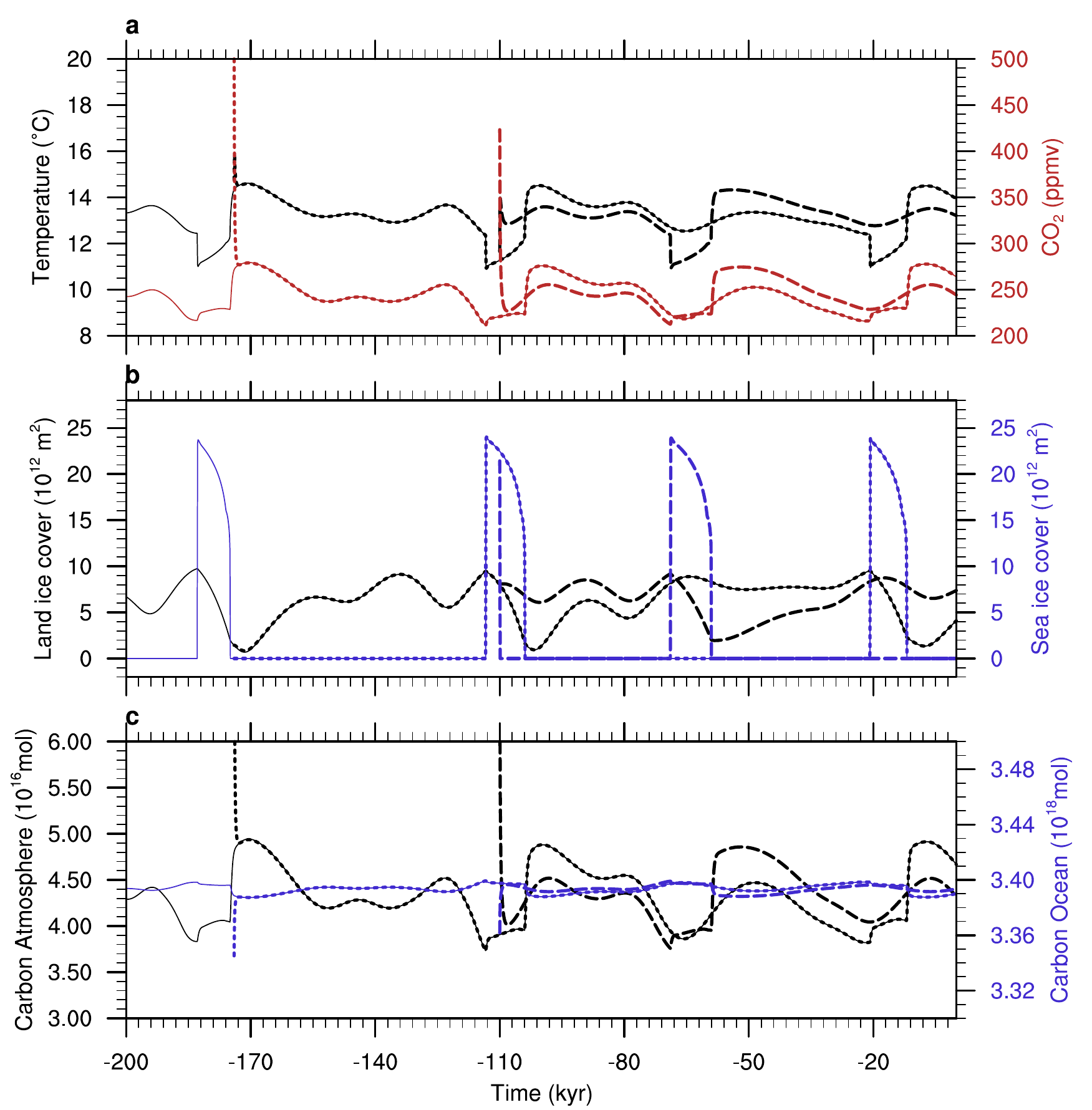}
\caption{Perturbed glacial cycles of the box model, shown are time series of 100-year averages. The perturbation consists of doubling the atmospheric CO$_2$ initially, while compensating for the added carbon in the ocean model. Perturbations are applied at two instances: during an interglacial (dotted lines) and during a glacial with extensive sea-ice cover (dashed lines). 
(a) Simulated global mean surface temperature $T$ (black lines) and atmospheric $\rm{CO}_2$ (red lines); 
(b) Land ice (black lines) and sea ice (blue lines) cover of the northern polar box; 
(c) Total carbon in the atmosphere (black lines) and the ocean (blue lines). 
}
\label{f:2xCpert_ts}
\end{figure}

\begin{figure}[ht!]
\center\includegraphics[width=7cm]{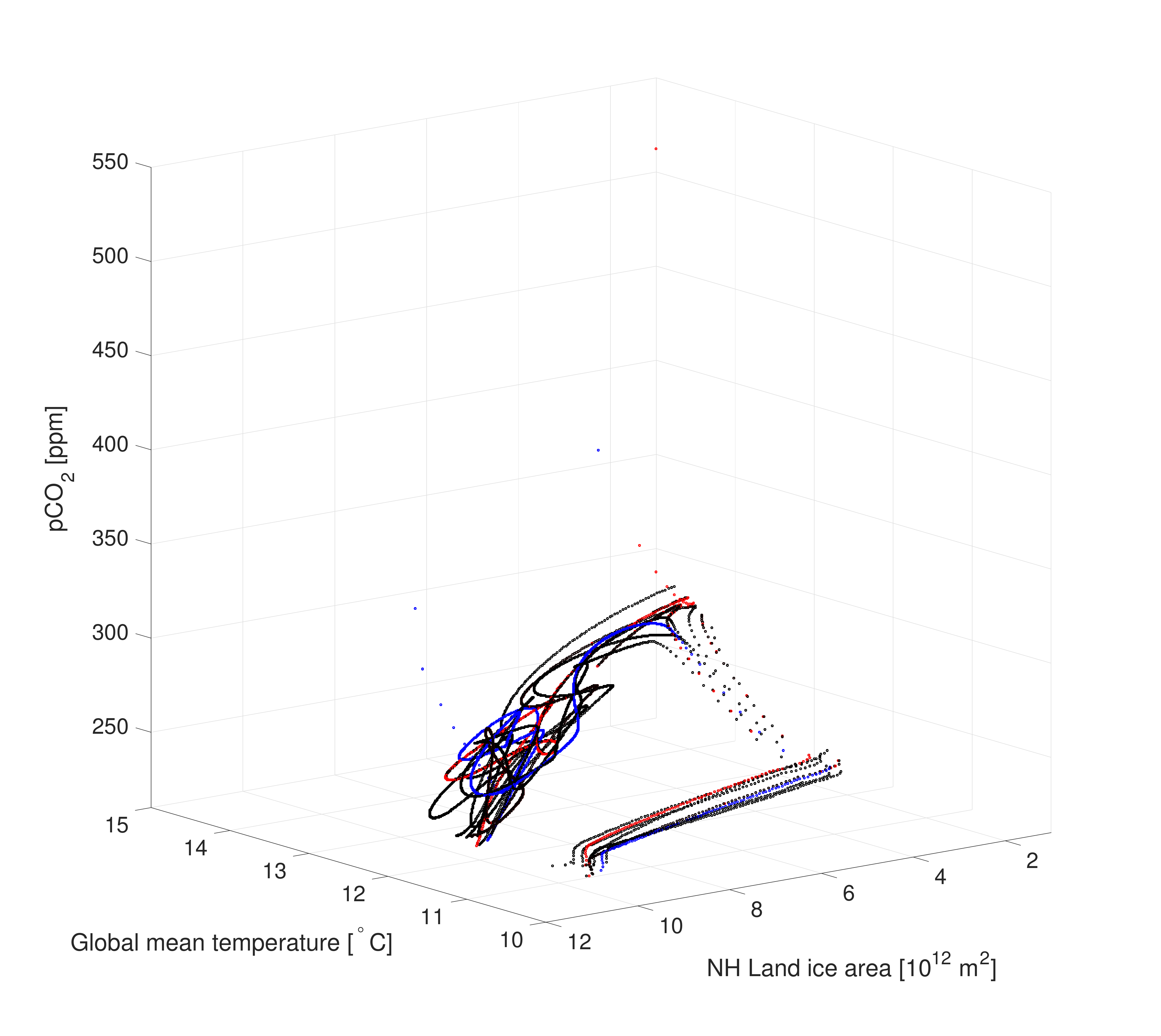}~\includegraphics[width=7cm]{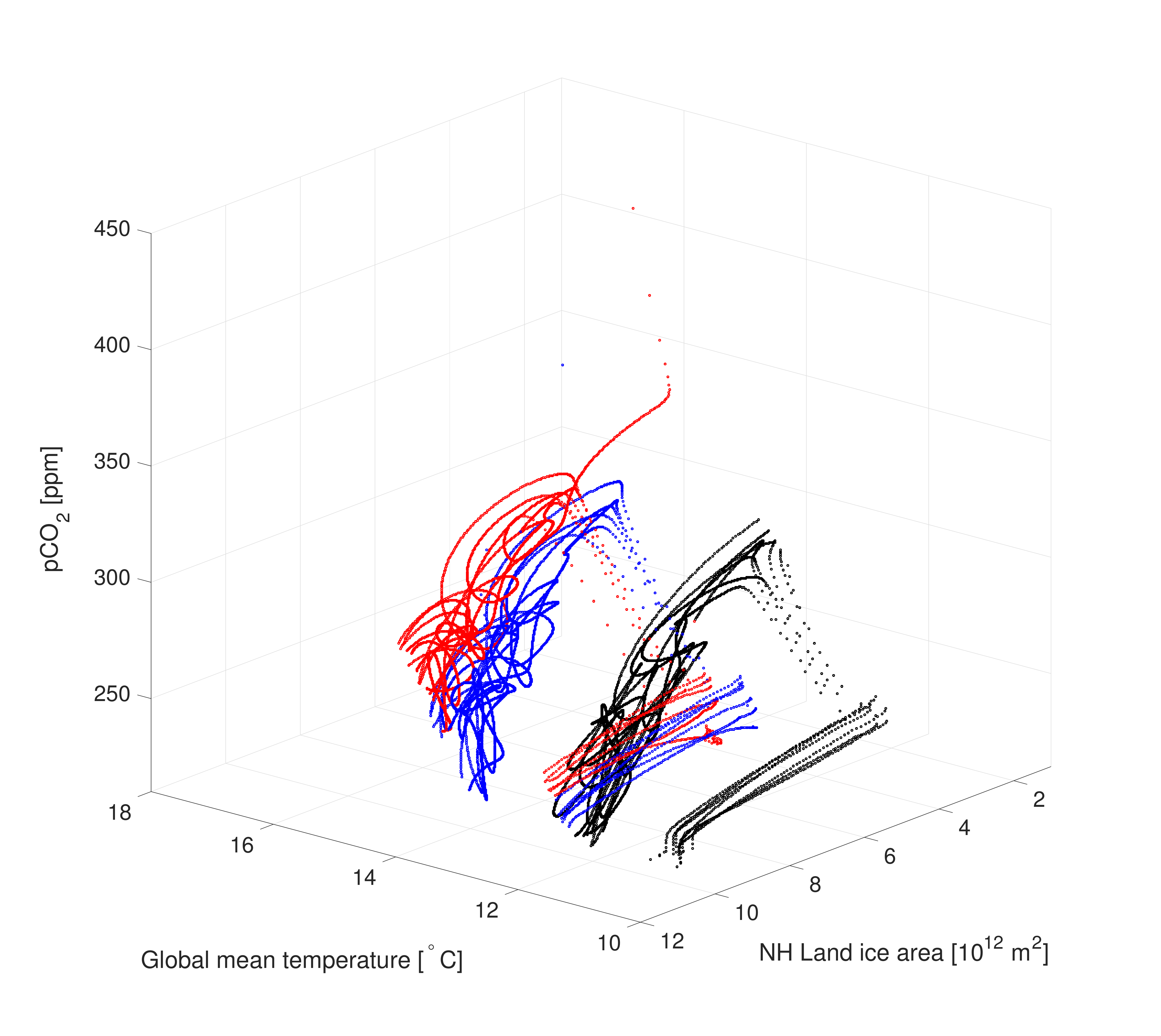}

\caption{Climate attractor of the climate model showing responses to doubling \coo\ from states in the $C$ regime (blue symbols) with and the $W$ regime (red symbols). Black symbols represent the unperturbed attractor as shown in Fig.~\ref{f:glacialcycles_p}a. We apply two different types of perturbations: {\bf left panel} The total amount of carbon in the model system (ocean and atmosphere) is conserved. When doubling the \coo concentration in the atmosphere, the same amount of \coo\ is removed from the ocean. After an inital transient, the model returns to the same attractor as schematically illustrated in Fig.~\ref{f:schematicTR}b; {\bf right panel} The atmospheric \coo\ is doubled without compensating in the ocean, meaning that extra carbon is added to the model system. In this case, the perturbed simulations return to an attractor that has a similar shape, but is shifted to higher \coo\ levels and global mean temperatures (see also Fig.~\ref{f:schematicTR}d). 
}
\label{f:2xCpert_attr}
\end{figure}

From the 500kyr model time series shown in Fig.~\ref{f:glacialcycles} we chose two initial conditions, one in the $W$ regime and one in the $C$ regime with extensive sea ice. The \coo\ is doubled initially in the atmosphere and the extra \coo\ added by the doubling is uniformly subtracted from the ocean boxes, in order to conserve the total amount of carbon in the model system. Note that the atmospheric \coo\ in this model is purely determined by the biological pump in the oceans \citep{gildor2002}. Both perturbed initial conditions are run till present day (time 0), time series are shown in Fig.~\ref{f:2xCpert_ts} together with the unperturbed time series (as in Fig.~\ref{f:glacialcycles}). The time series starting from the $W$ state quickly returns to the same temperature, \coo, sea- and land-ice time series (dotted lines in Fig.~\ref{f:2xCpert_ts}), where glacial inceptions and deglaciations occur at exactly the same time as in the unperturbed simulation (thin solid lines). In contrast, the perturbed time series starting from the $C$ state  (dashed lines) does not return to the same unperturbed time series; the sea ice present initially is melted by the initial warming, and the system undergoes a transition to the $W$ regime. While on the long term, the variation in temperature, \coo, sea ice and land ice covers the same range as in the unperturbed simulation, the timing of glacial inceptions and deglaciations is different from the unperturbed simulation. This suggests that the applied perturbation is indeed small enough such that the system returns to the same attractor as can be seen in the left panel of Fig.~\ref{f:2xCpert_attr}. When the initial condition lies in the $C$ regime, the perturbation induces a regime switch, after which the same attractor is explored, but on a different trajectory. 

We have also applied a modified perturbation, where the total amount of carbon is not conserved; atmospheric \coo\ is doubled, while the oceanic \coo\ is unchanged. In this case, the result of the perturbation is to shift the attractor towards higher temperatures, higher atmospheric \coo\ and slightly different amounts of land ice, as shown in the right panel of Fig.~\ref{f:2xCpert_attr}. This type of perturbation might reflect more realistically the present-day climate change situation assuming that the carbon injected into the system originates from a geological reservoir. However, while the attractor remains very similar in shape in this case but is shifted in phase space, other components of the model system such as the land ice might need adaptations of their parameters. The situation reflects the one depicted in Fig.~\ref{f:schematicTR}d and we will not further discuss the response to this type of perturbation but instead focus on the situation, where the climate system returns to the same attractor after the perturbation (Fig.~\ref{f:2xCpert_attr} left panel).

\begin{figure}[ht!]
\center\includegraphics[width=9cm]{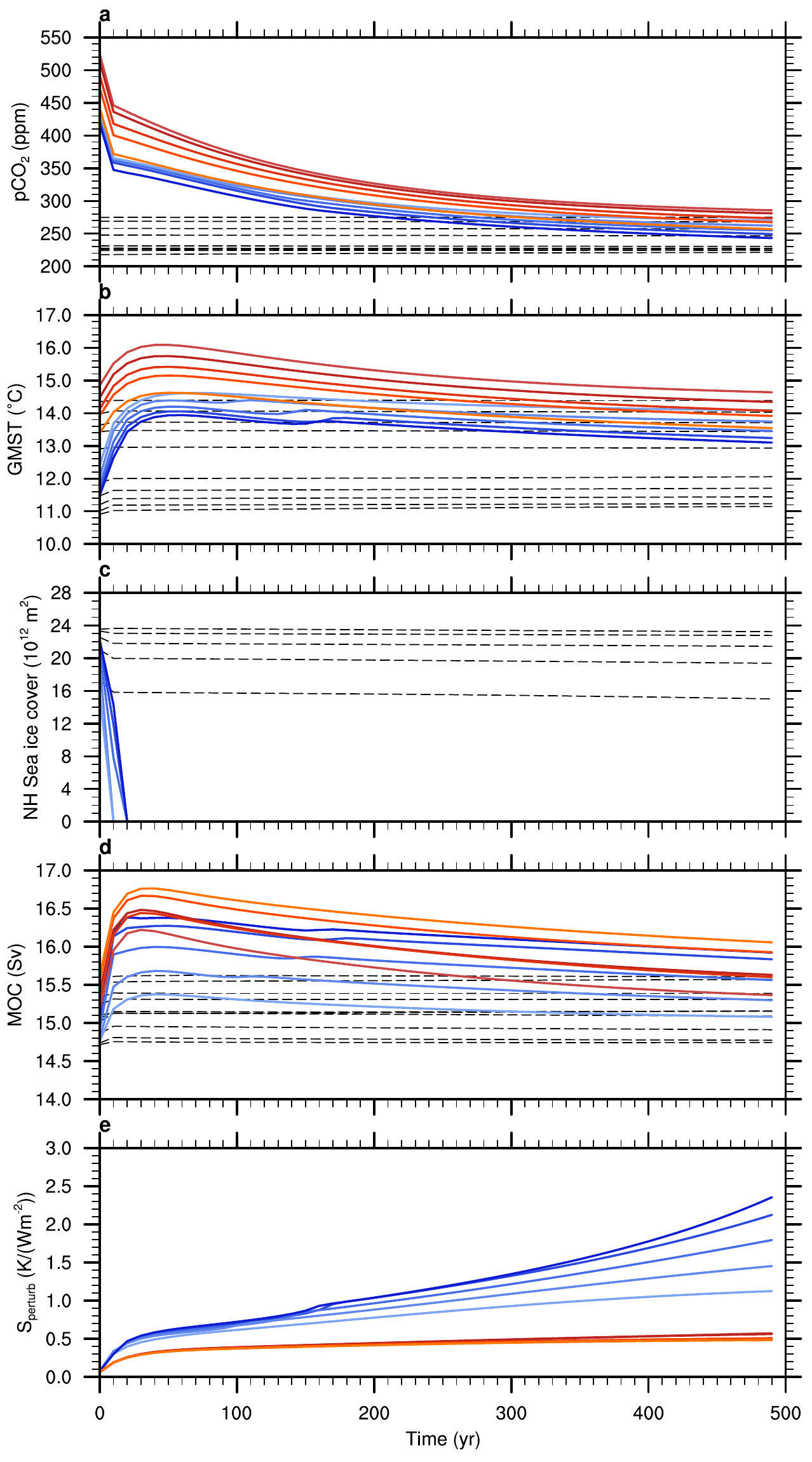}
\caption{Climate sensitivity $S_{perturb}$ from perturbation experiments with dynamical \coo. Shown are time series of experiments, where \coo\ is doubled initially and free to evolve until year 500 (coloured lines) along with the control experiments, where \coo\ is not doubled initially (black dashed lines). The ensemble starts from 250 initial conditions taken from the glacial-interglacial time series (500 kyr, shown in Fig.~\ref{f:glacialcycles}a). In this figure we show 10 ensemble members, the blue lines have sea ice initially, while the red lines have no sea ice and darker red indicates warmer initial temperature. Climate sensitivity is determined following eq.~\ref{e:S_perturb}. 
(a) Atmospheric \coo;
(b) Global mean surface temperature;
(c) Northern hemisphere sea ice fraction;
(d) Strength of the ocean meridional overturning circulation;
(e) $S_{perturb}$.}
\label{f:2xCpert_time}
\end{figure}

Starting from 250 different initial conditions chosen along the 500 kyr time series shown in Fig.~\ref{f:glacialcycles}a (one initial condition every 2000 years), the model is integrated for 500 years to give control runs of temperature $T^{cntrl(i)}(t)$ and radiative forcing time series $R^{cntrl(i)}_{[CO_2]}(t)+R^{cntrl(i)}_{[LI]}(t)$, respectively, where the index $i =  1,...,250$ denotes the initial condition. The initial \coo\ concentration $pCO_2^0$ in these simulations varies between 210 -- 290 ppm, while the global mean temperature varies between 10.8 -- 14.9$^\circ$C. A second set of simulations is performed, where the initial value of the \coo\ is doubled and then the model is integrated for 500 years, giving $T^{pert(i)}(t)$ and $R^{pert(i)}_{[CO_2]}(t)+R^{pert(i)}_{[LI]}(t)$.  A (time-dependent) climate sensitivity is then determined from
\be
S_{perturb}^{(i)}(t) = \frac{\Delta T^{(i)}(t)}{\Delta R^{(i)}(t)} = \frac{T^{pert(i)}(t)-T^{cntrl(i)}(t)}{R^{pert(i)}_{[CO_2]}(t)+R^{pert(i)}_{[LI]}(t)-R^{cntrl(i)}_{[CO_2]}(t)-R^{cntrl(i)}_{[LI]}(t)}
\label{e:S_perturb}
\ee
Fig.~\ref{f:2xCpert_time} shows time series of \coo, temperature, Northern Hemisphere sea ice cover, the ocean meridional overturning circulation strength and $S_{perturb}$ for a few of the ensemble members (both control and perturbed experiments). Clearly, the different time scales in the system become evident; the global mean atmospheric temperature reacts quickly to the elevated \coo-level, and for those initial states that have sea ice, the sea ice melts within 10-20 years. As the \coo\ is dynamic in these simulations, the increased \coo-gradient between ocean and atmosphere leads to a rather fast initial reduction in atmospheric \coo\ (timescale of $\sim 10$ years \citep{Gildor2002}), which then keeps decreasing on a longer time scale. After 500 years, temperature and \coo\ are almost back to their original values if the initial condition was within the $W$ regime. However, the initial conditions within the $C$ regime involve a regime shift and do not return to the same temperature and \coo-level within 500 years. The strength of the meridional overturning circulation in the ocean weakly responds to the \coo\ perturbation on a slower timescale as can be seen in Fig.~\ref{f:2xCpert_time}d. The time-dependent climate sensitivity $S_{perturb}$ is shown in Fig.~\ref{f:2xCpert_time}e; here the different behaviour of the $C$ and $W$ states becomes particularly evident: while the response to the perturbation of the $W$ states (red lines) seem to approach an 'equilibrium' value with some spread, increasing in time, the $C$ states (blue lines) produce a wide range of responses depending on where the attractor is met after the perturbation. 

\begin{figure}[ht!]
\center\includegraphics[width=10cm]{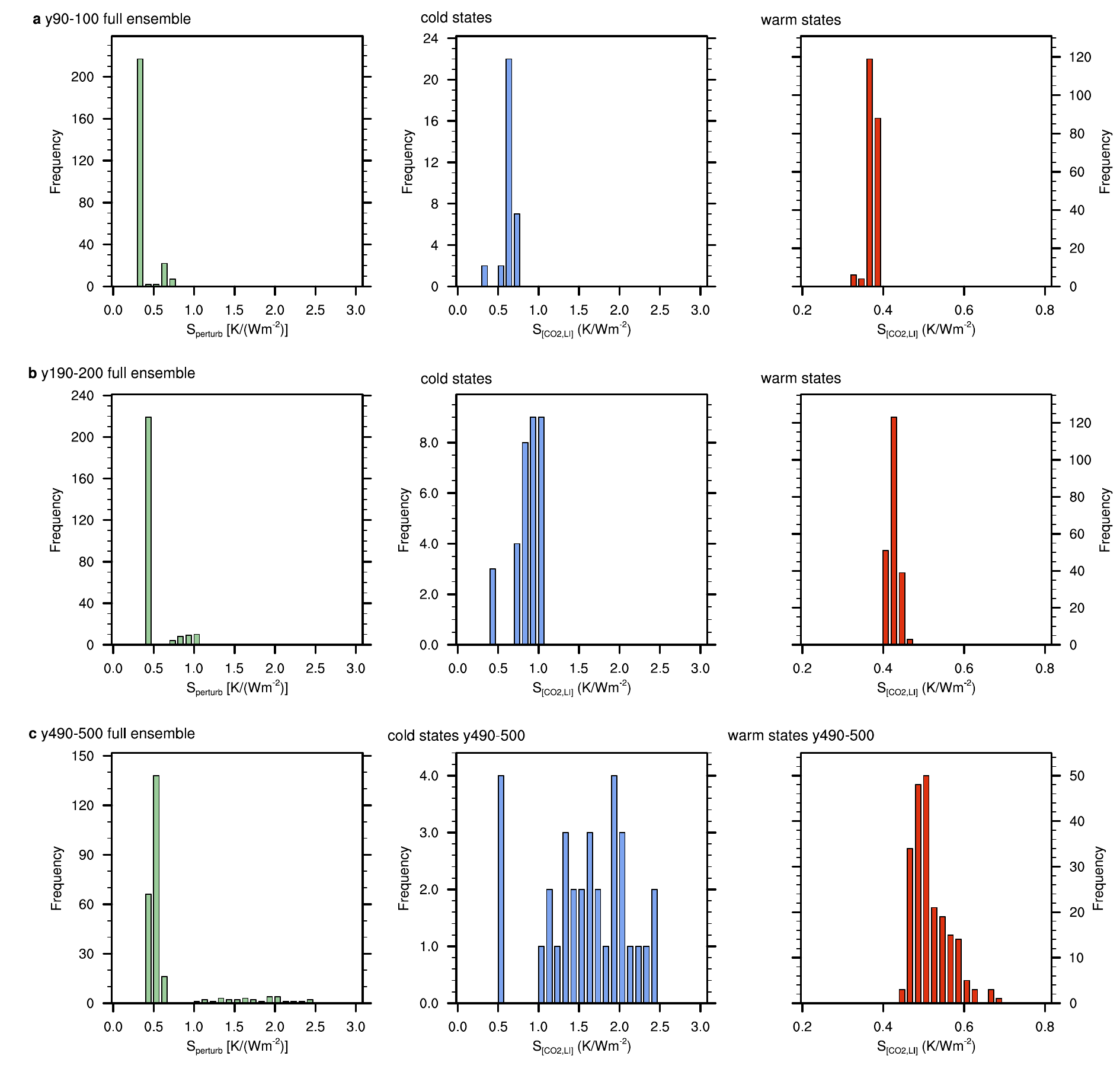}
\caption{Climate sensitivity $S_{perturb}$ from perturbation experiments with dynamical \coo. Shown distributions of $S_{perturb}$ (cf. eq.~\ref{e:S_perturb}) at different times after the perturbation. The ensemble starts from 250 initial conditions taken from the glacial-interglacial time series (500 kyr, shown in Fig.~\ref{f:glacialcycles}a). The right panel in each plot shows the full ensemble, the middle panel shows only those initial states that have sea ice (classified as $C$), and the right panel shows the warm initial states without sea ice (classified as $W$).
(a) $S_{perturb}$ after 100 years; 
(b) $S_{perturb}$ after 200 years; 
(c) $S_{perturb}$ after 500 years.
}
\label{f:2xCpert_hist}
\end{figure}

Snapshots of distributions of $S_{perturb}$ are shown in Fig.~\ref{f:2xCpert_hist} for 100, 200 and 500 years after the perturbation. A fast-process equilibrium should be expected after about 100-200 years, however, the spread in $S_{perturb}$ also increases with time, in particular for the $C$ regime. $S_{perturb}$ of the $W$ regime is similar to $S_{[CO_2,LI]}^{WW}$ after 100 and 200 years, but further spread out after 500 years. On the other hand, $S_{perturb}$ of the $C$ states resembles $S_{[CO2,LI]}^{CW}$ already after 200 years, and afterwards spreads out even further. In this ensemble, $S_{[CO2,LI]}^{CC}$ is in fact never observed, because the perturbation always induces a $C-W$ transition.

\section{Conclusions}
\label{s:concl}

In this paper, we have considered climate sensitivity as a local property of a climate attractor, in particular it is a property of a projection of a measure of this attractor on the $(T,R)$ plane. This naturally leads to distributions of climate sensitivity for every radiative forcing, and if the attractor shows different regimes of special climate dynamics, state dependence of climate sensitivity can be explained in terms of regimes. We have explored this in a conceptual Earth System model with the aim to test how climate sensitivity derived from palaeoclimate records might be compared to model-derived counterparts. Conceptually, climate sensitivity is defined differently in these two situations; while palaeoclimate timeseries reflect trajectories on the climate attractor, in model simulations generally perturbations away from the attractor are applied. Moreover, climate models include only a limited amount of processes (usually the slower processes are fixed, as is the carbon cycle), which means that a different attractor may be explored by the models. 

Clearly, we cannot expect to get reliable quantitative conclusions about the distribution of ECS from the low order conceptual model used for this study. Many important processes in the climate system (such as the impact of $T$ on cloud formation) are absent from the model, which was constructed in \citep{Gildor2001a,Gildor2002} with the aim of explaining ice age pacing rather than the link between $T$ and \coo. Even those processes that are included are open to debate; for example, the sea ice cover changes in the model are about 1.5 times larger than suggested by proxy data \citep{Koehler2010}, while Northern Hemisphere land ice cover changes are smaller (Fig.~\ref{f:glacialcycles}b). Moreover, the climate sensitivity derived from the model is higher during glacial periods, because the fast sea ice-albedo feedback is stronger in those regimes. Proxy data suggest, however, higher climate sensitivity during warm periods \citep{Heydt2014,Koehler2015}, most likely because a combination of other fast feedbacks (such as water vapour, cloud feedbacks, etc.) may be stronger during warm climates. 

Nonetheless, even for this model, the presence of variability on a number of timescales and regimes within the attractor gives clear and we find non-trivial dependence of sensitivity on regime. This suggests that it could be useful to think of the unperturbed climate sensitivity (which can be determined from palaeoclimate data) as a property of the ``climate attractor''. For a perturbed system (we have considered instantaneously doubled \coo), which is the normal approach in climate models, this is still useful once an initial transient has decayed. This transient will depend in particular on ocean heat uptake, though also on carbon cycle and biosphere processes that act on time scales roughly equivalent with the forcing time scale. 
In the case of a regime shift (either natural or induced by perturbation) the spread in climate sensitivity become very large.  
If the climate system has more than one attractor, the perturbed system may clearly evolve to a completely different set of states than the original attractor -- a situation that does not occur in the climate model used here. In less extreme cases, we cannot rule out very long transients (associated with slow feedbacks) for some perturbations.

In most climate sensitivity studies, feedback processes are considered except those related to the carbon cycle. In the history of climate, those processes are active, however, on many different timescales. In our conceptual model, we have included the part of the carbon cycle that is related to the soft-tissue biological pump in the oceans and air-sea \coo\ exchange. The resulting \coo\ variations in the model's glacial-interglacial cycles are in the range of the observed glacial to interglacial \coo\ changes and amplify the glacial-interglacial cycle while they are not necessary to generate those cycles. Accordingly, when exploring climate sensitivity from perturbation experiments with the same model, we have instantaneously doubled \coo\ and kept the model's carbon cycle active. This procedure ensures that the perturbation experiments eventually return to the same attractor as the unperturbed system.  
Such perturbations  (illustrated in Fig.~\ref{f:schematicTR}b,c) are not normally applied in climate models used for climate predictions \citep{IPCC2013}, where climate sensitivity is derived from model simulations considering prescribed, non-dynamic atmospheric \coo. 

In our conceptual model, we have also examined climate sensitivities from a classical climate model perturbation (not shown); \coo\ is doubled within the first 30 years of the simulation and kept fixed afterwards for 200 years. In this case we find significantly lower sensitivities and smaller spread than for $S_{perturb}$ obtained from doubling \coo\ with dynamic \coo. This emphasises the importance of including dynamic carbon cycle processes into climate projections. Moreover, it supports the idea that the real observed climate response may indeed be larger than the model predicted one, because those models never will include all feedback processes in the climate system. 
Furthermore, the carbon cycle includes more timescales and processes than considered here in this simple model. For example, processes related to the ocean-seafloor system include carbonate compensation and silicate weathering, which act on much longer timescales and have been suggested to be responsible for a mean atmospheric lifetime of anthropogenic \coo\ of 30-35 kyr \citep{Archer2005}. 

When deriving climate sensitivity from palaeoclimate records it is important to take account of potential state dependence and different climate regimes before drawing conclusions on the ECS distribution that may be relevant for future climate evolution. For the conceptual model we consider, the long tail in the ECS distribution from the unperturbed (palaeoclimate) timeseries mostly results from the cross-comparison of states within different regimes ($CW/WC$). Similarly, the applied perturbation in the model always induced a regime transition and consequently large ECS values if the initial condition was in the $C$ regime, but not in the $W$ regime. 
In the context of our model, these high ECS values would not be relevant for the present climate continuing the current regime. On the other hand, if the present climate is in a regime that is susceptible to a regime shift (either natural, or due to anthropogenic 'perturbation') very large ECS values may be possible and indeed relevant. By studying data and models of warmer-than-present climates in the palaeorecord we may be able to achieve information on potentially warmer climate regimes existing for perturbed versions of the climate attractor. 

\newpage

\appendix

\section{Model equations}
\label{ap:model}

\subsection{Ocean and sea ice}
The ocean consists of 2 layers of 4 meridionally oriented boxes, where the polar boxes extend from 45$^\circ$ to the pole and the equatorial boxes from the equator to 45$^\circ$, with meridional lengths $L_1,L_2,L_3,L_4$ the same as the atmopsheric boxes. All tracers such as temperature $T$, salt $S$ and biogeochemical variables are averaged over the two equatorial boxes, such that in fact the dynamics is determined by only three meridional boxes. The two vertical layers have thicknesses $D_{upper}$ and $D_{lower}$, respectively. 
The ocean model dynamics includes a simple frictional horizontal momentum balance, is hydrostatic and mass-conserving:
\begin{eqnarray}
0 &=& - \frac{1}{\rho_0}\frac{\partial p}{\partial z} - \frac{g}{\rho_0}\rho\\
0 &=& -\frac{1}{\rho_0}\frac{\partial p}{\partial y}-r v\\
0 &=& \frac{\partial v}{\partial y} + \frac{\partial w}{\partial z}
\end{eqnarray}
Here, $(y,z)$ are the meridional and vertical coordinates and $(v,w)$ the corresponding flow velocities, respectively. $p$ is the pressure, $g$ the gravitational constant, $\rho_0$ a reference density and $r$ a friction coefficient. In each box, temperature $T$ and salinity $S$ determine the density via the full nonlinear equation of state as recommended by \citet{UNESCO}. Temperature and salinity are determined by the following balances:
\begin{eqnarray}
\frac{\partial T}{\partial t} +\frac{\partial (vT)}{\partial y} +\frac{\partial (wT)}{\partial z}
&=&  
K_h \frac{\partial T}{\partial y}+K_v \frac{\partial T}{\partial z}+Q^{atm}_T+Q^{sea-ice}_T\\
\frac{\partial S}{\partial t} +\frac{\partial (vS)}{\partial y} +\frac{\partial (wS)}{\partial z}
&=&  
K_h \frac{\partial S}{\partial y}+K_v \frac{\partial S}{\partial z}+Q^{atm}_S+Q^{sea-ice}_S+Q^{land-ice}_S
\end{eqnarray}
where $K_h$ and $K_v$  are horizontal and vertical diffusion coefficients, respectively.  As in \citet{Gildor2002}, the vertical mixing of any tracer $tr$ (e.g., temperature, salinity or ocean CO$_2$) in the southern polar box is dependent on the vertical stratification:
\begin{equation}
K_v^0(\sigma_{t_{deep}}-\sigma_{t_{surface}})^{-1}(tr_{deep}-tr_{surface}),
\end{equation}
where $(\sigma_{t_{deep}}-\sigma_{t_{surface}})\sim d\rho/dz$. In addition, upper and lower bounds of 280 and 1 Sv are imposed on the vertical mixing rates $K_v^0(\sigma_{t_{deep}}-\sigma_{t_{surface}})^{-1}$. Vertical mixing rates between the other surface and deep boxes are set constant, 0.25 Sv for the two equatorial boxes and 5 Sv for the northern polar box. The meridional overturning circulation is treated in the same way as in \citet{Gildor2002}, with the upwelling through the southern polar box set to a fixed value of 16 Sv and the downwelling through the northern polar box determined by the meridional density gradient between the northern equatorial and polar ocean boxes. 

The $Q$ terms in the above equations are fluxes from other components of the climate model: $Q^{atm}_T$ is the atmosphere-ocean heat flux due to sensible, latent and radiative fluxes:
\begin{equation}
Q^{atm}_T = \frac{\rho_0C_{pw}D_{upper}}{\tau}(\theta - T)(f_{ow}+f_{si}\frac{\gamma}{D_{sea-ice}}),
\end{equation}
where $C_{pw}$ is the heat capacity of water, $\theta$ the temperature of the atmospheric box above, $\gamma$ the insolation effect of a layer sea ice of thickness $D_{sea-ice}$. $f_{ow}$ and $f_{si}$ are the fractions of the ocean that are open water and sea ice covered, respectively, with $f_{ow} = 1-f_{si}$. The time scale $\tau$ is chosen such that the ocean heat transport into the northern polar atmopsheric box is about 2.3PW during interglacial periods as in \citet{Gildor2001a}. 
Precipitation $P$ and evaporation $E$ is converted into an equivalent salt flux:
\begin{equation}
Q^{atm}_S = -(P-E)S_0,
\end{equation}
with $S_0$ a reference salinity. Heat and salt fluxes due to sea ice formation or melting are formulated as:
\begin{eqnarray}
Q^{sea-ice}_T &=& \frac{\rho_0C_{pw}V_{ocean}}{\tau_{sea-ice}}(T^{sea-ice}-T),\\
Q^{sea-ice}_S &=& \frac{Q^{sea-ice}_T}{\rho_{sea-ice}L_f}S_0,
\end{eqnarray}
where $V_{ocean}$ is the volume of the ocean box, $T^{sea-ice}$ is the temperature threshold where sea ice forms, $L_f$ is the latent heat of fusion, $\rho_{sea-ice}$ the density of sea ice and $\tau_{sea-ice}$ is a short time scale to ensure that the ocean temperature remains close to the freezing temperature as long as sea ice is present. Sea ice is assumed to grow in area with an initial thickness of 3 and 1.5m in the northern and southern polar boxes, respectively, until the whole box is covered. 
The volume of sea ice in the polar surface boxes $V_{sea-ice}$ is given by:
\begin{equation}
\frac{dV_{sea-ice}}{dt} = \frac{Q^{sea-ice}_T}{\rho_{sea-ice} L_f}+P_{on-ice}.
\end{equation}
$P_{on-ice}$ is the amount of sea ice forming due to atmopsheric precipitation falling on the ocean area covered with sea ice.

\subsection{Atmosphere}

The atmospheric model follows that used in \citet{Gildor2002}, with 4 atmospheric boxes above the ocean boxes. The lower surface of each atmospheric box can be either land or ocean, and  both can be partly covered with (land or sea) ice. The box-averaged potential temperature is calculated from the energy balance of the box, balancing incoming solar radiation (with a box albedo determined from the relative fraction of each lower surface type in the box), outgoing longwave radiation at the top of the atmosphere, air-sea heat flux and meridional atmospheric heat transport. 
In each atmopsheric box, the temperature $\theta$ is determined by the difference between the heat flux at the top of the atmosphere $F_{top}$ and at the surface $F_{surface}$, 
following the  equation:
\begin{eqnarray}
\frac{\partial \theta}{\partial t}  &=&  \frac{2^{R/C_p}g}{P_0C_p}\left[(F_{top}-F_{surface})+(F^{out}_{merid}-F^{out}_{merid})\right]\\
&=& \frac{2^{R/C_p}g}{P_0C_p}\left[(H_{in}-H_{out}-Q^{atm}_{T})+(F^{in}_{merid}-F^{out}_{merid})\right],
\end{eqnarray}
where
\begin{eqnarray}
H_{in}&=&(1-\alpha_{surf})(1-\alpha_C)(1-q_{in}^{seaice})Q_{Solar}\\
H_{out}&=&\left(\varepsilon-\kappa\ln{\left(\frac{{\rm CO}_2}{{\rm CO}_{2,ref}}\right)}\right)\sigma_B \theta^4,
\end{eqnarray}
are the incoming and outgoing radiation terms at the top of the atmosphere, respectively. ($R$ is the gas constant for dry air, $C_p$ is the specific heat of the atmosphere at a constant pressure, $P_0$ a reference pressure, $\sigma_B$ the Stefan Boltzmann constant and $g$ the gravitational acceleration.) The incoming solar radiation $Q_{Solar}$ for each box is assumed to vary with season and due to orbital variations as in \citet{Gildor2000}. Furthermore, $Q_{Solar}$ is reduced by a constant cloud albedo term $\alpha_C$ and a part $q_{in}^{seaice}$ that is directly used to melt sea ice; where sea ice exists, 15\% of the incoming shortwave radiation is used to melt sea ice and does not enter the radiation balance of the atmosphere \citep{Gildor2002}. $\alpha_{surf}$ is the surface albedo of the box and is determined by the fraction of sea ice, land ice, land surface and ocean surface in that box:
\begin{equation}
\alpha_{surf}=f_L(1-f_{LI})\alpha_L+f_Lf_{LI}\alpha_{LI}+f_O(1-f_{SI})\alpha_O+f_Of_{SI}\alpha_{SI}
\label{e:albedo}
\end{equation}
Here, $f_L$, $f_{LI}$, $f_O$, $f_{SI}$ correspond to the fraction of land, land ice, ocean and sea ice, respectively, and $\alpha_L$, $\alpha_{LI}$, $\alpha_O$, $\alpha_{SI}$ to the corresponding albedos of each surface type. 
The outgoing radiation depends on a mean emissivity of the box $\varepsilon$ and a term depending on the atmospheric CO$_2$ concentration. Here $\kappa$ is chosen \citep{Gildor2002} such that a doubling of CO$_2$ will cause a radiative forcing of 4 Wm$^{-2}$. 
$F^{in}_{merid}-F^{out}_{merid}$ is the net heating due to meridional heat fluxes between the atmospheric boxes. Meridional heat transport between boxes is calculated as:
\begin{equation}
F_{merid} = K_\theta \nabla \theta,
\end{equation}
where the coefficient $K_\theta$ is chosen such that the meridional heat transport between the two northern boxes is about 2.2 PW during interglacial periods \citep{Gildor2001a}. 
No net heat flux is assumed over land and land ice, therefore $F_{surface}$ includes only the ocean-atmosphere heat exchange.

The meridional moisture transport $F_{Mq}$ between the atmospheric boxes is parameterised as:
\begin{equation}
F_{Mq} =  K_{Mq}|\nabla\theta|q,
\end{equation}
where $q$ is the humidity of the box. A constant relative humidity is assumed, with the saturation humidity at temperature $\theta$ calculated from an approximate Clausius-Clayperon equation:
\begin{equation}
q=0.7\cdot A \cdot e^{B/\theta}.
\end{equation} 
Over land ice in the polar boxes, another source of precipitation is the local evaporation of that part of the ocean box that is not covered by sea ice, with flux:
\begin{equation}
F_q = K_q f_{ow} q.
\end{equation}
The total precipitation in each box is then given by
\begin{equation}
P-E = -\nabla\cdot(F_{Mq}+F_q).
\end{equation}
Precipitation falling over land or sea ice is assumed to turn into additional ice. 

\subsection{Land ice}

The equations for the land ice sheets follow those of \citet{Gildor2001a}, with the mass balance
\begin{equation}
\frac{dV_{ice-sheet}}{dt} = LI_{source}-LI{sink}.
\end{equation}
The source term $L_{source}$ depends on the amount of precipitation falling over existing ice (or falling on the 0.3 poleward area of the box even if there is no glacier there):
\begin{equation}
LI_{source} = \frac{max\{0.3L_{area},LI_{area}\}}{box_{area}}(P-E),
\end{equation}
where $L_{area}$ is the land area in the box, $LI_{area}$ the ice sheet area and $box_{area}$ the total area of the box. 

The ice sheet can shrink as a consequence of ablation. The ablation term is assumed a constant $C_{LI}$ \citep{Gildor2001a} plus a modulation by the summer Milankovitch forcing \citep{Gildor2000}:
\begin{equation}
LI_{sink} = C_{LI} + \gamma_{LI}(Solar_{June} - Solar_{ave, June}),
\end{equation}
where $Solar_{June} - Solar_{ave, June}$ is the anomaly in summer insolation in this box relative to the average over the past 1 Myr. Southern hemisphere ice sheets are assumed constant. 

\subsection{Biogeochemistry}

In the ocean boxes additional tracers are advected for total CO$_2$ ($\Sigma CO_2$), alkalinity ($A_T$) and phosphate PO$_4$. These are used to calculate atmopsheric $pCO_2$, see \citet{Gildor2002}. The equations for the three biogeochemistry variables $Bio$ in each ocean box follow:
\begin{equation}
\frac{\partial Bio}{\partial t} +\frac{\partial (vBio)}{\partial y} +\frac{\partial (wBio)}{\partial z}
=
K_h \frac{\partial Bio}{\partial y}+K_v \frac{\partial Bio}{\partial z}+S_{Bio},
\end{equation}
with additional source/sink terms $S_{Bio}$ for these variables in the surface boxes:
\begin{eqnarray}
S_{\Sigma CO_2} &=& -R_{C}\times EP - RR\times EP+PV([CO_{2,a}]-[CO_{2,o}])\\
S_{A_T} &=& -2\times RR \times EP + R_N\times EP\\
S_{PO_4} &=& -EP,
\end{eqnarray}
and in the deep boxes below:
\begin{eqnarray}
S_{\Sigma CO_2} &=& R_{C}\times EP + RR\times EP\\
S_{A_T} &=& 2\times RR \times EP - R_N\times EP\\
S_{PO_4} &=& EP. 
\end{eqnarray}
$EP$ and $RR$ stand for export production and rain ratio, respectively, and $R_C$, $R_N$ for the ratio $P:C$ and $P:N$ in particulate organic matter, respectively. $[CO_{2,a}]$ is the saturation concentration with regard to the partial pressure of CO$_2$ in the atmosphere, and $[CO_{2,o}]$ is the CO$_2$ concentration in the ocean. The flux of \coo\ between ocean and atmosphere $F_{CO_2} = PV\left([CO_{2,a}]-[CO_{2,o}]\right)A_{open water}$ is linearly related to the $pCO_2$ difference between the atmosphere and the surface ocean via a constant piston velocity $PV$, giving a time scale of about 10 years for this gas exchange. For more details on the biogeochemistry module, see \citet{Gildor2002}. 

\section{Climate sensitivity in the model}
\label{ap:cs}

Climate sensitivity is determined from the energy balance of the Earth. For the conceptual model \citep{Gildor2001a}, we can explicitly write the energy balance of the atmosphere and extract the different contributions to climate sensitivity. 
Averaged over all atmospheric boxes of the model the global mean temperature $T = \sum_{i=1}^{4}{\frac{area_i}{area}\theta_i}$ 
is determined by the difference between the heat flux at the top of the atmosphere $F_{top}$ and at the surface $F_{surface}$ (see previous section), where $area_i$, ($i =1,...,4$) is the surface area of the 4 boxes and $area$ is the total surface area of the Earth. 

To access the contributions of the different forcings and feedbacks to the radiation balance, we split the global mean radiation terms into the different components due to solar radiation ($R_{[ins]}$), land ice ($R_{[LI]}$), sea ice ($R_{[SI]}$), outgoing longwave radiation ($R_{[OLW]}$), CO$_2$ concentration ($R_{[{\rm CO}_2]}$) and the radiation at the Earth's surface ($R_{[surf]}$):
\begin{equation}
\frac{\partial T}{\partial t} =   \frac{2^{R/C_p}g}{P_0C_p} \left[R_{[ins]}+R_{[LI]}+R_{[SI]}+R_{[OLW]}+R_{[{\rm CO}_2]}+R_{[surf]}\right]
\label{e:rad}
\end{equation}
The different contributions to the radiation balance can be expressed as:
\begin{eqnarray}
R_{[ins]}&=&(1-\alpha_C)Q_{solar}\\
R_{[LI]}&=&R_{[ins]}\sum_i \frac{area_i}{area}(f_L^i(1-f_{LI}^i)\alpha_L+f_L^if_{LI}^i\alpha_{LI})(q_{in}^{seaice}-1)
\label{e:rad_li}\\
R_{[SI]}&=&-R_{[ins]}\sum_i \frac{area_i}{area}[q_{in}^{seaice}+(1-q_{in}^{seaice})(f_O^i(1-f_{SI}^i)\alpha_O+f_O^if_{SI}^i\alpha_{SI})]\label{e:rad_si}\\
R_{[OLW]}&=&-\sum_i \frac{area_i}{area}\varepsilon_{i}\sigma_B \theta_i^4\label{e:rad_olw}\\
R_{[{\rm CO}_2]}&=&\sum_i \frac{area_i}{area}\kappa\ln{\frac{pCO_2}{pCO_{2,ref}}}\sigma_B \theta_i^4\label{e:rad_co2}\\
R_{[surf]} &=&-\sum_i \frac{area_i}{area}Q_{oa}^i\label{e:rad_surf}.
\end{eqnarray}
 
When comparing two equilibrium climate states with global mean temperatures 
$T_1$ and $T_2$ (and $\Delta T = T_2 - T_1$), the radiation balance Eq.\ref{e:rad} reads:
 \begin{equation}
 0 = \Delta R_{[ins]}+\Delta R_{[LI]}+\Delta R_{[SI]}+\Delta R_{[OLW]}+\Delta R_{[{\rm CO}_2]}+\Delta R_{[surf]}.
 \end{equation}
 As we consider constant solar radiation and no changes in cloud albedo, $\Delta R_{[ins]} =0$, and, when we put all the forcing or slow feedbacks on the left hand side and all fast feedback processes on the right hand side, we obtain: 
\begin{equation}
 \Delta R_{[{\rm CO}_2]}+\Delta R_{[LI]} = -\Delta R_{[OLW]}-\Delta R_{[SI]}-\Delta R_{[surf]}.
 \end{equation}
 This finally leads to the expressions for the specific climate sensitivities
 \begin{eqnarray}
S_{[{\rm CO}_2]}&=&\frac{\Delta T}{\Delta R_{[{\rm CO}_2]}} = \frac{-\Delta T}{\Delta R_{[OLW]}+\Delta R_{[SI]}+\Delta R_{[surf]}+\Delta R_{[LI]}}\\
S_{[{\rm CO}_2,LI]}&=&\frac{\Delta T}{\Delta R_{[{\rm CO}_2]}+\Delta R_{[LI]}}= \frac{-\Delta T}{\Delta R_{[OLW]}+\Delta R_{[SI]}+\Delta R_{[surf]}}
\label{e:specS}\\
S_{[{\rm CO}_2,LI,SI]}&=&\frac{\Delta T}{\Delta R_{[{\rm CO}_2]}+\Delta R_{[LI]}+\Delta R_{[SI]}}= \frac{-\Delta T}{\Delta R_{[OLW]}+\Delta R_{[surf]}}.
\end{eqnarray}
The last expression should approximate the sensitivity without feedbacks (i.e. only Planck feedback), $S_0 = (-4\varepsilon\sigma_BT^3)^{-1}\simeq 0.3$~\KWmm. In the model there is, however, one more radiation term due to the atmosphere-ocean heat exchange ($\Delta R_{surf}$), which acts on fast to intermediate time scales. Therefore, $S_{[{\rm CO}_2,LI,SI]}$ still slightly deviates from the Planck sensitivity.

\section*{Acknowledgements}

This work was carried out under the program of the Netherlands Earth System Science Centre (NESSC), financially supported by the Ministry of Education, Culture and Science (OCW) in the Netherlands. AH thanks CliMathNet (sponsored by EPSRC) for travel support to meetings that facilitated this work. We thank the Lorentz Center in Leiden for organising a ``Workshop on Climate Variability: from Data and Models to Decisions'' in 2014 where these ideas were first discussed, and the EU ITN ``CRITICS'' for providing a further opportunity to discuss this research.

\bibliographystyle{plainnat}
\setcitestyle{authoryear,open={(},close={)},semicolon}

\newpage

\end{document}